\documentclass[sigconf,nonacm]{acmart}
\RequirePackage[frozencache]{minted}

\AtBeginDocument{%
  }

\setcopyright{acmcopyright}
\copyrightyear{2018}
\acmYear{2018}
\acmDOI{XXXXXXX.XXXXXXX}

\acmConference[Conference acronym 'XX]{Make sure to enter the correct
  conference title from your rights confirmation emai}{June 03--05,
  2018}{Woodstock, NY}
\acmPrice{15.00}
\acmISBN{978-1-4503-XXXX-X/18/06}

\graphicspath{{./}}

\usepackage{xcolor}
\usepackage{soul}

\definecolor{sqhlcolor}{HTML}{c7edf4}
\newcommand{\sq}[1]{{\sethlcolor{sqhlcolor}\hl{#1}}}

\usepackage{mdframed}
\global\mdfdefinestyle{notetoreaderstyle}{backgroundcolor=yellow!50}

\newcommand{\nop}[1]{}

\usepackage{xcolor}
\usepackage{framed}

\definecolor{mygray}{gray}{0.85}

\renewenvironment{quote}{
  
  \MakeFramed{\advance\hsize-\width \FrameRestore}
  \noindent\begin{minipage}{\dimexpr\linewidth-1em\relax}
}{
  \end{minipage}
  \endMakeFramed
}

\begin{document}

\title{ShellVis: Sandboxed Live Programming for Shell Scripts}

\author{Joshua Horowitz}
\email{joshuah@alum.mit.edu}
\orcid{0000-0002-5154-9277}
\affiliation{%
  \institution{University of Washington}
  \city{Seattle}
  \country{United States}}

\author{Jeffrey Heer}
\email{jheer@uw.edu}
\orcid{0000-0002-6175-1655}
\affiliation{%
  \institution{University of Washington}
  \city{Seattle}
  \country{United States}}


\begin{abstract}
Live programming provides visibility to programmers by running and tracing programs as they are edited.
However, for programs with potentially harmful side effects, liveness can turn mistakes into disasters.
We propose enabling live programming in environments with side effects via sandboxing: confining effects to a simulation of the true environment.
We apply \emph{sandboxed live programming} in the challenging context of shell scripting: a ubiquitous and powerful---yet notoriously opaque and error-prone---tool.
\emph{ShellVis} provides line-by-line feedback on a shell script's run-time behavior, with file operations sandboxed via a safe overlay of the file system.
A qualitative user evaluation finds ShellVis to be helpful to participants, replacing tedious existing practices and instilling confidence.
Participant responses also reveal areas for future research, particularly bridging the gulf of execution alongside the gulf of evaluation.
ShellVis serves as a case study of how sandboxing can bring live-programming techniques into the many real-world programming contexts where side effects are important.
\end{abstract}


\begin{teaserfigure}
\includegraphics[width=\linewidth]{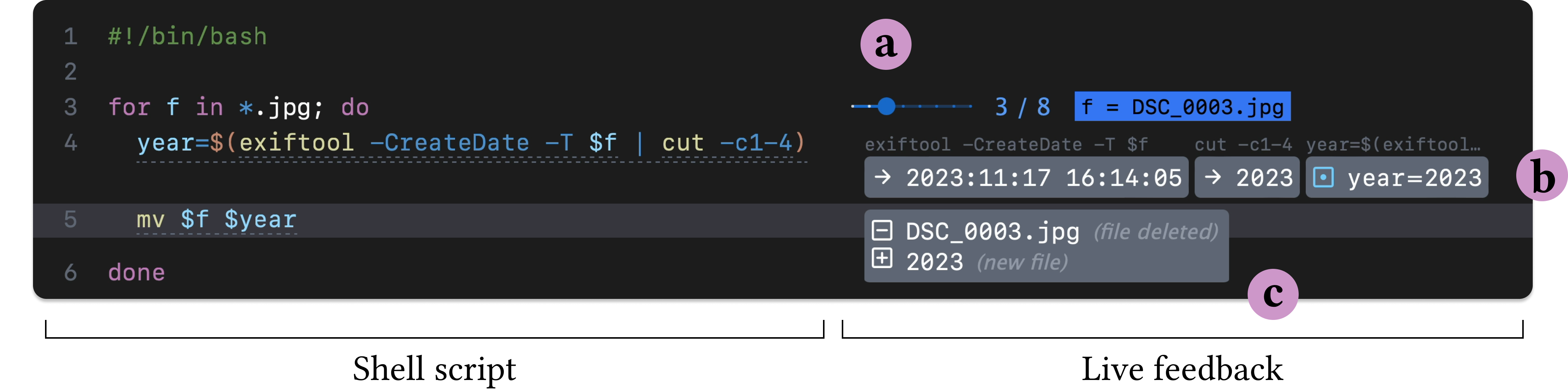}\caption{A ShellVis user writes a script to sort photos into folders by year. The script's source is shown on the left. On the right, focus+context visualizations convey script outputs. A loop slider (a) indexes iterations of the \texttt{for} loop, while abbreviated bubbles (b) show command output and effects on shell variables. An expanded bubble for the current focus (c) shows file system effects. The script contains a critical mistake, but ShellVis ran it in a sandbox so the user's photos are unharmed.}\Description{A screenshot of the ShellVis interface. It shows an interface with shell-script code on the left-hand side and output bubbles & widgets on the right-hand side, aligned with lines of code. An "mv" command produces an output bubble reporting that a image file has been deleted and a file named "2023" has been created.}
\label{fig:teaser}
\end{teaserfigure}

\maketitle

\section{Introduction}
\label{sec:introduction}

While computer use has generally moved from the command line to GUIs, the shell remains a essential site of activity for users who find value in its precision and power. The shell is also a gateway guiding end users into programming \cite{Nardi1993-ln}. Everything a user types at a command line is written in a programming language like bash \cite{bash}. By packaging a list of commands into a text-file program, a \emph{shell script}, a user can document a process and make it re-usable. Starting with this straight-line automation, further levels of programmability open up: parameterizing scripts with command-line arguments, iterating through files with \texttt{for} loops, and so on.

Shell scripts have a much broader reach, both actual and potential, than just system administrators and software engineers. For instance, shell scripts have been adopted by many practitioners that work with data, including data journalists \cite{data-journalists}, scientists \cite{scientists}, and sociologists \cite{sociologists}. Indeed, since nearly all computer users work with files and programs---the main characters of shell scripts---scripts have the potential to be useful to nearly everyone. We envision a world where shell scripts are approachable tools, amplifying the power of diverse computer users.

In practice, the picture is not so rosy. Many obstacles lie in the way of diverse users wielding shell script super-powers. While these include the learning challenges shared by all programming languages, shell scripts present unique challenges which make them ``powerful, ubiquitous, and reviled'' \cite{doi:10.1145/3458336.3465294} even among experienced programmers. These include idiosyncratic syntax and unfamiliar semantics, in a ``fast-paced high-stakes'' environment where ``a single typo could erase entire hard drives'' \cite{doi:10.1145/3458336.3465294}.

We hypothesize that these downsides can be significantly mitigated with an interactive programming environment. In particular, we draw upon the techniques of \emph{live programming} \cite{doi:10.22152/programming-journal.org/2019/3/9} -- running a program as the programmer edits it and providing detailed feedback on what their in-progress program does. With live programming, a programmer can see the actual data running though their program and plan their next step in response. They can quickly detect problems and survey program behavior to diagnose root causes.

While live programming has proven successful in contexts like spreadsheets and computational notebooks, it faces new challenges when applied to shell scripts. Shell scripts are typically written not (only) to compute numbers or make graphs, but to perform \emph{side effects}, such as modifying files, which are typically difficult to undo. How can liveness, which relies upon running a program over and over again as it is modified, be made to work in an environment where running the program might be dangerous?

To explore these questions, we contribute \textbf{ShellVis}, a live programming system for bash shell scripts made possible through \emph{sandboxing}. As a script is edited in a conventional text editor, ShellVis runs it, presenting the programmer with a fine-grained, navigable display of its run-time behavior, including text written to output streams, changes to shell variables, and modifications to the file system. To make sure every run of the program starts from the same ``clean slate'', and to protect the user from irreversible mistakes, ShellVis runs the script in a virtual file system which simulates changes to the user's real file system without actually modifying it.

Our goals with ShellVis are to help programmers understand programs (both their own \& others') and write programs more effectively. To assess this, we ran a user study in which seven participants with 1-20 years of shell experience tried out ShellVis for the first time, performing four shell-scripting tasks. Results were encouraging, with participants unanimously expressing that ShellVis was helpful, comfortable to use, and desirable for their own work. From participants, we learned how the feedback \& sandboxing provided by ShellVis could replace tedious existing practices. Critically, participants reported that ShellVis made them feel safer and more confident as they worked with shell scripts, a quality we believe is especially important in end-user-programming and learning contexts. Finally, we identified ways that ShellVis does not yet meet users' needs, from helping them figure out what code to write in the first place, to sandboxing more kinds of effects, to better integrating live feedback with existing editors.

In summary, this paper makes the following contributions:

\begin{itemize}
\item Identifying \emph{sandboxed live programming} as a way to provide fine-grained feedback to programmers working in contexts, like shell scripting, centered around side effects.
\item ShellVis, a prototype sandboxed live-programming environment for shell-scripting that uses overlay file systems and visualization techniques to provide safe, fine-grained feedback to programmers.
\item A user evaluation of ShellVis demonstrating that programmers found the combination of live-programming and sandboxing to be helpful, while revealing both the efficacy of ShellVis's approach and needed future directions for sandboxed live programming to reach full effectiveness.
\end{itemize}

\noindent{}Together, these contributions suggest there is untapped potential for fine-grained live programming to support real-world programming contexts, if side-effects can be appropriately sandboxed.

\section{Background and related work}
\label{sec:background-and-related-work}

\subsection{Shell Scripts}
\label{sec:shell-scripts}

Unix-based operating systems (including Linux and Mac OS) provide a command interpreter called the \emph{shell} \cite{doi:10.1145/361011.361061, doi:10.1007/3-540-09745-7_2}. The shell lets users perform one-off commands to start and stop processes, manipulate the file system, and more. On top of these atomic operations, the shell provides ways to compose operations together: the celebrated inter-process pipe \cite{doi:10.22152/programming-journal.org/2023/7/13, doi:10.1145/361011.361061} as well as basic programming constructs like variables and loops. With these scripting constructs, the shell becomes a tool for automation.

On GUI-based computers, graphical mechanisms have largely replaced the shell: users launch applications from icons and drag files between windows representing directories. But even in this setting, the shell maintains unique advantages, including scriptability. While a user can drag a single file in a file manager, they cannot record this action for later re-use, or generalize it to a broader pattern to run in a loop. Graphical automation is an active field of research, and can be found in modern OSes in places like Shortcuts \cite{shortcuts}, but many users still reach for the shell for its precise control and its established ecosystem of command-line tools.

Unfortunately, ``{[}s{]}hell scripts are hard even for experts'' \cite{Greenberg2021ReportOT}. We have been unable to find formal studies in the literature specifically focused on challenges with shell scripts, but general dissatisfaction with the shell is broadly acknowledged. Greenberg et al.~\cite{doi:10.1145/3458336.3465296} do not hesitate to call it ``a confusing and dangerous object of disgust.'' Participants in their HotOS 2021 panel \cite{Greenberg2021ReportOT} proposed a number of contributory factors, including opaque shell syntax, a lack of visibility into shell state, and dangerous, irreversible operations. Testimony from participants in our own study support these hypotheses (\S\ref{sec:rq1}). ShellVis seeks to address these obstacles through \emph{sandboxed live programming}.

\subsection{Programming Environments}
\label{sec:within-code}

It is helpful to position ShellVis as a \emph{within-code live-programming environment}, and to place this category in relation to the neighboring concepts of \emph{REPL}, \emph{notebook}, and \emph{debugger}.

The Unix command line is a REPL (read-eval-print loop) for the shell language. REPLs provide fast feedback, enabling experimentation and iterative development. However, going from running one-off commands in a REPL to building a durable program is not always easy, and REPLs in many other languages have been supplanted in practice. For instance, while Python can be run in a REPL, interactive use of Python has shifted strongly towards \emph{notebooks} like Jupyter \cite{Kluyver2016JupyterN}. A notebook can reproduce the append-only pattern of a REPL, but it doesn't need to. Notebook cells are mutable, and can be revised and re-run. In this way, notebooks combine the document-like persistence of a text-file script with the immediacy and visibility of a REPL. The success of notebooks raises the question (briefly alluded to by Greenberg et al.~\cite{doi:10.1145/3458336.3465294}) as to whether shell scripts could similarly benefit from a hybrid interface, partway between REPL and traditional script editor.

ShellVis is such a hybrid interface. Rather than taking the form of a notebook, it takes the form of a \emph{within-code live-programming environment} (WCE): an environment that ``augments {[}a{]} textual-code editor\ldots{} with in-context displays showing run-time behavior'' \cite{lrc}. While notebooks achieve liveness with a cell-based structure, WCEs like Projection Boxes \cite{doi:10.1145/3313831.3376494} provide live feedback within a traditional textual programming environment.\footnote{For more examples see Horowitz \& Heer~\cite{lrc}.}

WCEs have some advantages over notebooks. They co-exist seamlessly with the corpus and ecosystem of a language. Existing programs can be loaded into a WCE and existing tools can be used alongside it. As a text-based environment, a WCE benefits from broad support for plain text, from powerful editors to revision control systems like git. Finally, WCEs can bring live feedback ``deeper'' into a program than is afforded by a notebook, such as into loop and function bodies \cite{doi:10.1145/3586183.3606733}. We chose to build ShellVis as a WCE for these advantages.

Like WCEs, \emph{debuggers} provide visibility into run-time behavior. However, they are designed for a fundamentally different purpose. While debuggers are used to investigate a program after something has gone wrong, WCEs provide feedback to support writing a program in the first place. Debuggers rarely support this latter task. Traditional single-step debuggers only provide delicate, keyhole visibility into a program, rather than always-on feedback. Omniscient debuggers which ``make it possible to navigate backwards in time within a program execution trace'' \cite{doi:10.1145/1297105.1297067} are closer in design and spirit to WCEs, but are still not designed to provide feedback during iterative program development.

\subsection{Live Programming with Side Effects}
\label{sec:side-effects}

Live programming relies on re-running programs (or parts thereof) in response to edits so programmers can see the effects of their edits. For this reason, live programming has historically been most at home in pure functional environments like spreadsheets, where re-runs are safe and predictable.\footnote{Pure functional programming is based on producing new values from old values without modifying (``mutating'') values in place. Spreadsheet formulas follow the pure functional discipline. Pure functional programs are easy to re-run; running the same program with the same inputs produces the same outputs. As an example of how pure functional architecture benefits liveness: Cells in Jupyter notebooks cannot be reliably re-run in response to upstream changes because information flows between cells via hard-to-track mutations. In contrast, \emph{reactive} notebooks like Observable follow the spreadsheet execution model \cite{observable}, producing a more live experience.}

What happens when we remove this constraint and make imperative systems live? Numerous live environments (including Projection Boxes \cite{doi:10.1145/3313831.3376494} and Seymour \cite{seymour}) have provided visibility into programs that mutate local state: state contained \emph{within} a program which can be cleanly re-initialized on each execution. ShellVis takes on a larger challenge: creating a live programming experience for shell scripts that interact with the outside world through \emph{side effects}.

There are two reasons why side effects cause problems for live programming. First, the execution of a script in a live-programming environment is intended to be \emph{speculative}, showing what the program \emph{would} do if run in a particular starting world, so the programmer can iterate until the program does what they want. If this program caused real side effects as it re-ran, the programmer would be working on constantly shifting ground. We need to make programs with side effects \emph{repeatable}. Secondly, by moving to a regime where a program is executed automatically and repeatedly as it is edited rather than at deliberately chosen points, live programming increases the chance that a side effect will cause a disastrous, irreversible result. We need to make programs with side effects \emph{safe}.

The need for live programming to grapple with external side-effects has recently been recognized by Sulír et al.~\cite{doi:10.1145/3593434.3593501}, who concluded from their study of Java source code that ``neglecting I/O is not a viable option for tool designers''. Yet, ``studies about this phenomenon are rare'' and ``I/O is rarely taken into account when designing {[}live programming{]} tools.''
In order to make a reactive, live programming experience possible, ShellVis takes the approach of sandboxing access to the file system. We are not aware of any other live-programming projects that sandbox or otherwise contain their external effects, as ShellVis does.

\begin{figure*}[t]
\centering
\includegraphics[width=\linewidth]{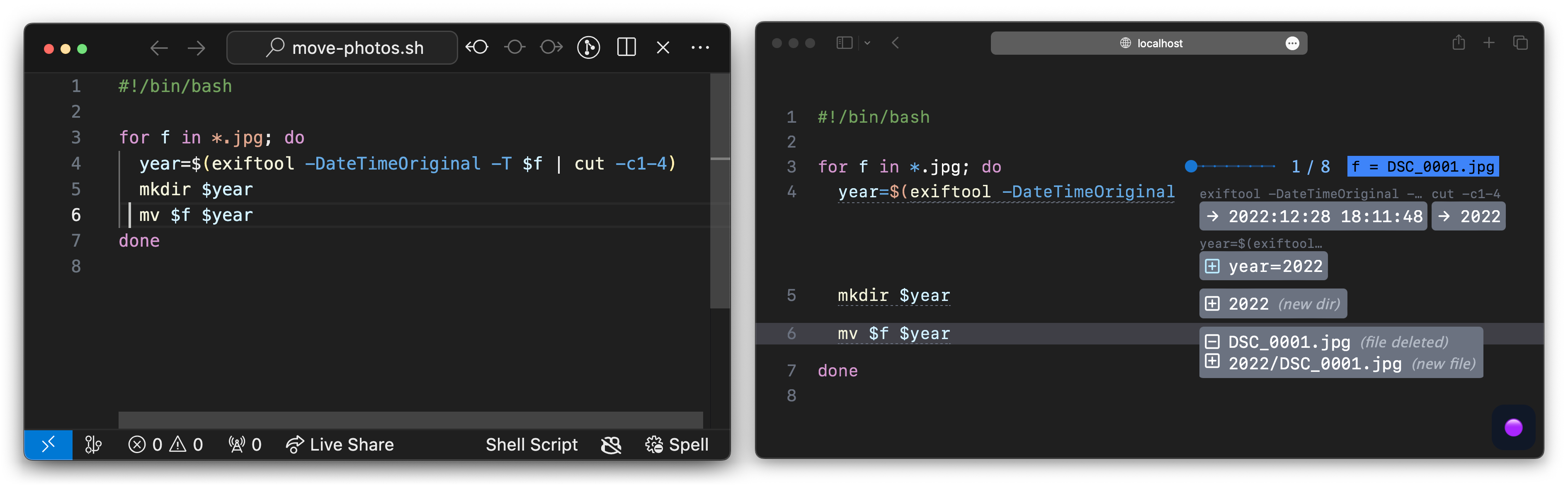}\caption{ShellVis in use on a user's desktop. A text editor (Visual Studio Code) is open on the left and the ShellVis visualizer is open in a web browser on the right. The visualizer's focus is linked to the text cursor position in the editor.}\Description{The code in the text editor matches code shown on the left-hand side of the ShellVis visualizer, though the visualizer also shows output bubbles and widgets. A line in the visualizer is highlighted, matching the position of the cursor in the editor}
\label{fig:split-screen}
\end{figure*}

\subsection{Tools for Shell Scripting}
\label{sec:shell-tools}

The tool most relevant to this work is \texttt{try}, a command-line program for Linux which ``lets you run a command and inspect its effects before changing your live system'' \cite{try}. To use \texttt{try}, a user takes a command they want to run at the command line and prepends \texttt{try} to it. \texttt{try} then runs the target command in a partial sandbox which simulates writes to the file system without actually performing them. It reports to the user what files have been modified by the command, and offers the option of ``committing'' the changes.

Sandboxing in ShellVis is directly inspired by \texttt{try}. Specifically, we borrow \texttt{try}'s use of an ``overlay file system'' to protect the original file system and detect file system modifications to report to the user (\S\ref{sec:sandboxing}).
ShellVis's unique insight is to employ sandboxing in the context of a larger live programming environment. ShellVis provides not just a top-level summary of what a script does to the file system (as \texttt{try} does), but fine-grained information on what individual commands in a script do. ShellVis complements information on file system modifications with other streams of information, including command output and changes to shell variables (\S\ref{sec:design}).

Live programming tools for authoring shell pipelines include Ultimate Plumber \cite{up} and Pipecut \cite{pipecut}. ShellVis generalizes these tools, visualizing not just pipelines but entire scripts. ShellVis also sandboxes side effects, while these prior tools do not.

Several debuggers exist for shell languages. BASHDB is a single-step debugger for Bash \cite{bashdb}. The Shtepper is a research program stepper for POSIX-compatible shell scripts built on the Smoosh executable semantics \cite{doi:10.1145/3371111}. Farther from Unix, Windows PowerShell has traditional single-step debugging \cite{powershell}. Shell debuggers do not seem to be used much, at least with the Unix shell. For visibility and debugging, the tool of choice is \texttt{echo}, the shell's ``print'' statement. Some programmers may also use features like \texttt{set -x} and \texttt{set -v} which tell the shell to log all commands as they run.

Many other tools are available for working around the quirks and corner cases of the shell. ShellCheck is a popular static analysis tool that detects dangerous patterns in scripts \cite{shellcheck}. NoFAQ suggests fixes to shell commands based on error messages \cite{doi:10.1145/3106237.3106241}. explainshell breaks complex shell commands into pieces and provides documentation from \texttt{man} pages \cite{explainshell}. Finally, LLM-powered code assistants have been used for shell scripting as they have for every other domain of programming. We consider all these tools to be complementary to ShellVis, and discuss code assistants in particular more in \S\ref{sec:gulfs}.

\subsection{Summary}
\label{sec:summary}

Tools providing visibility into the run-time behavior of shell scripts are rare and inadequate. ShellVis fills this gap with a \emph{within-code live programming environment} which shows the user as much information as possible about the behavior of an under-development script. A key enabling technology for this is \emph{sandboxing} the script's effects to make them repeatable and safe. While sandboxing shell scripts has been explored before in \texttt{try}, sandboxing has not yet been combined with fine-grained feedback. We believe this combination provides a powerful and unexplored synergy.

\section{Sandboxed live programming in ShellVis}
\label{sec:sandboxed-live-programming-in-shellvis}

Like most ``within-code live-programming environments'' (e.g., Projection Boxes \cite{doi:10.1145/3313831.3376494}), ShellVis uses a four-step feedback loop:

\begin{enumerate}
\item \textbf{Editing:} Provide the user with a means to edit a program.
\item \textbf{Running:} At some cadence, automatic or user-defined, run the program.
\item \textbf{Tracing:} Monitor the program's execution to extract fine-grained information.
\item \textbf{Viewing:} Report tracing information to the user in a helpful way.
\end{enumerate}

\noindent{}Before jumping into a demonstration, we provide give a quick overview of how ShellVis approaches these four steps.

\textbf{Editing:} ShellVis lets users edit code in the text editor of their choice rather than requiring that they move to a custom editor. Existing editors are polished, familiar, and entrenched. Any tool demanding programmers leave behind their favorite text editor faces an uphill battle. We also chose this path simply to focus on our core objective: the provision of run-time feedback.

\textbf{Running:} The running stage is where ShellVis introduces the essential component of \emph{sandboxing}. Shell scripts can have many different effects, including file system access, network access, interaction with other processes (including launching or killing processes), access to devices, and assorted system calls. We chose in this project to focus on sandboxing file system access.

The file system is a natural starting point for sandboxing. It is a central site of action and coordination in modern operating systems. Explicit file management is a primary use-case for shell scripts, and the file system is used as backing state for many infrastructures (like git). We leave sandboxing of other effects to future work, but we discuss a broad range of ways they might be handled in \S\ref{sec:sandboxing-more}.

\textbf{Tracing:} The machinery ShellVis uses to sandbox file system access can also be leveraged to tell us what effects each command in a script has on the file system: what files are added, removed, or modified. This tracing of external effects provides essential information to programmers to make sure their scripts are doing what they expect. We augment this with additional trace information on process output, as well as changes in shell state such as variables and working directory.

\textbf{Viewing:} Presenting the large amount of traced information to the user in a helpful way is a significant challenge. ShellVis manages this complexity with a focus+context ``fisheye view'' \cite{doi:10.1145/22627.22342, doi:10.1145/1124772.1124921} that adjusts the level of detail based on the user's current focus, and with interactions for surveying loop iterations.

\section{Usage Scenario}
\label{sec:shellvis-by-example}

Figure~\ref{fig:split-screen} shows the layout of a user's screen while using ShellVis. A text editor (Visual Studio Code) is open on the left, displaying a finished shell script. On the right, the ShellVis visualizer is open inside a web browser. It displays the same script, together with output annotations inside gray ``bubbles.'' As the user edits the script on the left, the display on the right updates.

\begin{figure*}[p]
\centering
\includegraphics[width=0.94\linewidth]{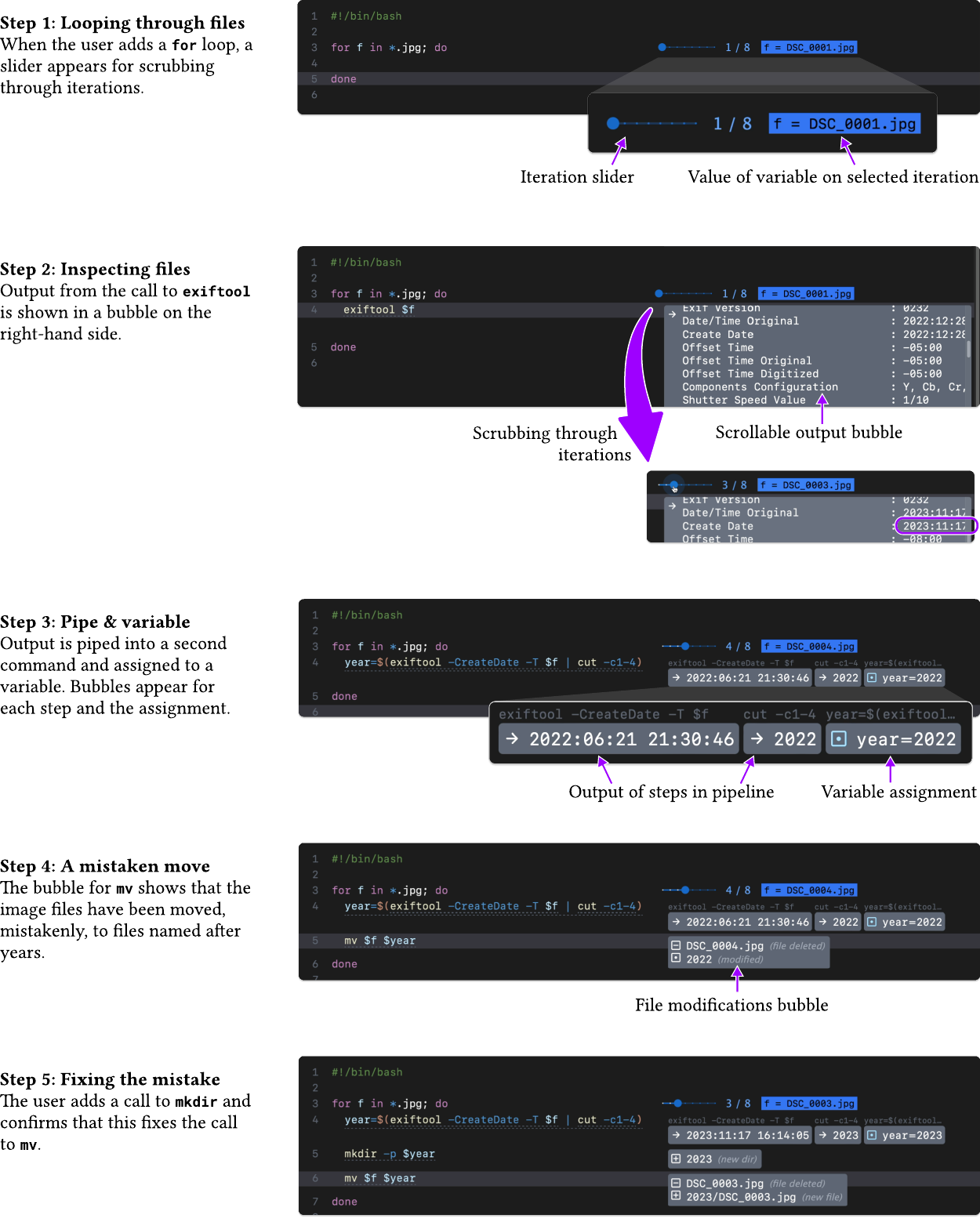}\caption{Using ShellVis to sort photos into folders by year. Screenshots show the visualizer interface (the right half of Figure~\ref{fig:split-screen}).}\Description{Full-page image split into five steps, each showing a screenshot of the ShellVis interface. Step 1: The script is a loop with an empty body. A slider appears on the right-hand side, representing the iterations of the loop. It says "1 / 8", and gives the current value of the loop variable with the text "f = DSC_0001.jpg". Step 2: The loop now contains a call to exiftool. It gets a large output bubble showing (a piece of) a large output. One row of the output gives a "Create Date" which looks good. A small second screenshot shows how sliding the slider to "3 / 8" produces different output, with a different value for "Create Date". Step 3: The loop now contains the code "year=$(exiftool -CreateDate -T $f | cut -c1-4)". ShellVis visualizes this with a bubble for exiftool output, a bubble for cut output, and a bubble for the variable assignment which reads "year=2022". Step 4: The loop now contains a call to "mv $f $year". Its bubble shows information about added and deleted files: an image file has been deleted and a file named "2023" has been created. Step 5: A call to "mkdir -p $year" has been added between the "year=..." line and the "mv" line. Its output bubble nowsays a file named "2023/DSC_003.jpg" has been created.}
\label{fig:demo}
\end{figure*}

The purpose of the script shown is to sort a directory of JPEG photos into subdirectories based on the year each photo was taken. Let's examine how ShellVis assists the user through the steps of making this script, with Figure~\ref{fig:demo} as a visual aid.

\textbf{Step 1: Looping through files.} The user wants to perform an action for each JPEG image, so they start with a \texttt{for} loop. They type out a loop with an empty body and press Ctrl-S to save. When ShellVis detects that the script has changed, it automatically re-runs it and displays a visualization of its execution to the user. In this case, a slider appears to the right of the line with \texttt{for}, reporting that the loop has run 8 times, and on its first iteration the variable \texttt{f} has taken on the value ``DSC\_0001.jpg''. Scrubbing the slider left and right, the user can see what other values this variable takes on. This is the sort of information that might ordinarily be obtained by adding an \texttt{echo \$f} in the loop, but ShellVis displays information like this automatically and in-context. Once the loop's body is no longer empty, this slider will also control which iteration is used as context for visualization of the body's execution.

\textbf{Step 2: Inspecting files.} Now the user wants to figure out what year these photos were taken. They add a call to \texttt{exiftool \$f}, which prints out the file's EXIF metadata. This output is displayed in a standard-output bubble. In this case, the output is large and the bubble lets the user scroll its contents. They scroll past several irrelevant date fields to find one that looks correct. Scrubbing through iterations of the for loop, they confirm that the target date field changes as expected.

\textbf{Step 3: Pipe \& variable.} To extract the year from the \texttt{exiftool} output, the programmer performs a number of steps. They pass some flags into \texttt{exiftool} to return just the ``Create Date'' field. Then they pipe this into \texttt{cut -c1-4} to return just the first four characters of this date -- the year. Finally, they assign this result to the shell variable \texttt{year}. As they perform each of these changes, they save the file to see its effects in the visualizer. By the time they have finished this step, there are three bubbles visible: one for \texttt{exiftool}, one for \texttt{cut}, and one for the variable assignment. This script would not ordinarily print the outputs of \texttt{exiftool} and \texttt{cut}, as these are intercepted by ShellVis for display. The variable assignment bubble displays a different icon meaning ``variable changed.''

\textbf{Step 4: A mistaken move.} Now the user tries to move the photos into directories named after years, with \texttt{mv \$f \$year}. The bubble on the right shows that the original file at \texttt{DSC\_0004.jpg} is gone, and there is now a file at \texttt{2022}. This is a mistake! Since directories for years do not yet exist, \texttt{mv} interprets its arguments as asking to rename the photo to a file named after the year. In fact, a photo was moved to \texttt{2022} in an earlier iteration of the loop, and in this fourth iteration, it is being destructively overwritten with a second photo taken in 2022. Ordinarily, this mistake would be either inconvenient (if backups were made) or disastrous (if not). Fortunately, ShellVis sandboxes file system access: the user's original files have not been touched and all writes to files are taking place in a simulated layer. The user can simply recognize the mistake and iterate on their script to fix it. Every execution of the script starts from the same state.

\textbf{Step 5: Fixing the mistake.} To fix their mistake, the user adds \texttt{mkdir \$year} before \texttt{mv}. With this change, \texttt{mv}'s output bubble shows the move is performed successfully. The script is complete and the user can run it in their terminal.

\section{Interface Design}
\label{sec:design}

In this section, we describe ShellVis's design in more detail, focusing on how we augmented code with feedback. The following section goes into more detail on implementation, including sandboxing.

\subsection{Scope}
\label{sec:scope}

ShellVis is intended to bring live programming into shell programmers' workflows in an incremental way while remaining compatible with existing practices.\footnote{This goal of ``incremental adoptability'' fits under the ``Sociability'' dimension in the ``Technical Dimensions of Programming Systems'' classification \cite{doi:10.22152/programming-journal.org/2023/7/13}.} This led to several specific design decisions. One was to build ShellVis as a \emph{within-code live programming environment} (\S\ref{sec:within-code}). Here we discuss two other, related decisions.

First, we decided that ShellVis should support standard POSIX-family scripts (like bash and zsh), rather than using an alternative language or switching away from plain-text programming entirely. This decision was made with some reluctance. Compared to modern languages, the shell's syntax and semantics are idiosyncratic, and this seems to cause real problems for programmers. Many appealing alternatives are available.\footnote{These include Xonsh \cite{xonsh}, YSH \cite{ysh}, Nushell \cite{nushell}, Fish \cite{fish}, and Murex \cite{murex}.} However, POSIX-family shell scripts are both technically and culturally entrenched. By building on them, ShellVis can be used with the enormous corpus of existing scripts and in the many contexts where shell scripts are necessary. Furthermore, from a research point of view, building a live-programming system on traditional shell scripts allows us to better understand the contribution of a live-programming interface without the complicating factor of a change in language, and it helps us maintain an approachable scope.

Second, we decided to make ShellVis work alongside any existing editor, rather than making our own editor or embedding ShellVis deeply inside a particular editor. For prototyping purposes, we could have built a custom editor on top of an open-source library like CodeMirror \cite{codemirror}. However, we felt that this would yield few benefits, and would have to be abandoned if we ever wanted to distribute ShellVis for actual use: existing editors are just too good, and programmers are just too attached to them. Embedding ShellVis deeply inside of an existing editor is an appealing idea. However, it would limit the reach of ShellVis to users that prefer that editor. Also, the extension APIs of editors are rarely general enough to support the particular visualizations and interactions we wanted ShellVis to have. This is a general problem: Projection Boxes \cite{doi:10.1145/3313831.3376494} could not be implemented without forking the Visual Studio Code codebase, making distribution much more difficult.

\subsection{Fisheye Information Bubbles}
\label{sec:bubbles}

\begin{figure}[t]
\centering
\includegraphics[width=0.90\linewidth]{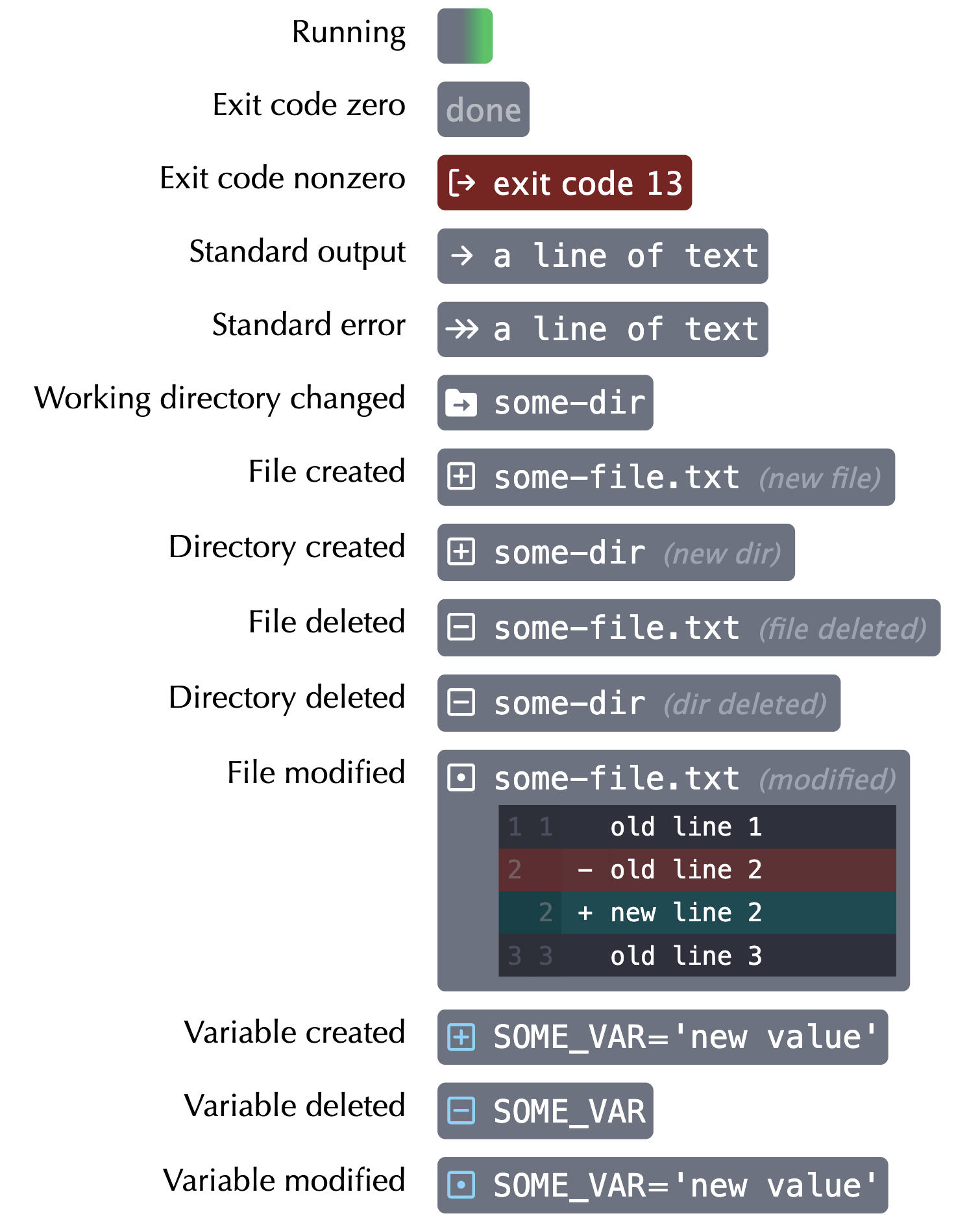}\caption{Annotations supported by ShellVis}\Description{A listing of labelled output bubbles. Running: a small empty bubble with a green gradient moving across it. Exit code zero: a bubble with the text "done". Exit code nonzero: a red bubble with an "exit" icon and the text "exit code 13". Standard output: a bubble with an arrow and the text "a line of text". Standard error: a bubble with a double arrow and the text "a line of text". Working directory changed: a bubble with a directory-arrow icon and the text "some-dir". File created: a bubble with a plus icon and the text "some-file.txt (new file)". Directory created: a bubble with a plus icon and the text "some-dir (new dir)". File deleted: a bubble with a minus icon and the text "some-file.txt (file deleted)". Directory deleted: a bubble with a minus icon and the text "some-dir (dir deleted)". File modified: a bubble with a dot icon, the text "some-file.txt (modified)", and a "diff" visualization showing changes to lines. Variable created: a bubble with a blue plus icon and the text "SOME_VAR='new value'". Variable deleted a bubble with a blue minus icon and the text "SOME_VAR". Variable modified: a bubble with a blue dot icon and the text "SOME_VAR='new value'".}
\label{fig:gallery}
\end{figure}

As ShellVis runs a shell script, it produces a program trace which it streams to the web browser visualizer. The visualizer displays the trace to the programmer, in the form of ``information bubbles'' attached to commands run by the script. Figure~\ref{fig:gallery} shows the annotation bubbles currently supported by ShellVis. These annotations are all generic, in the sense that they show information that can apply to all commands. We believe they encompass most forms of feedback that shell-script authors need to see to understand the behavior of their scripts, with the exception of feedback about effects that ShellVis does not currently sandbox, like network access.

\begin{figure}
\centering
\includegraphics[width=0.90\linewidth]{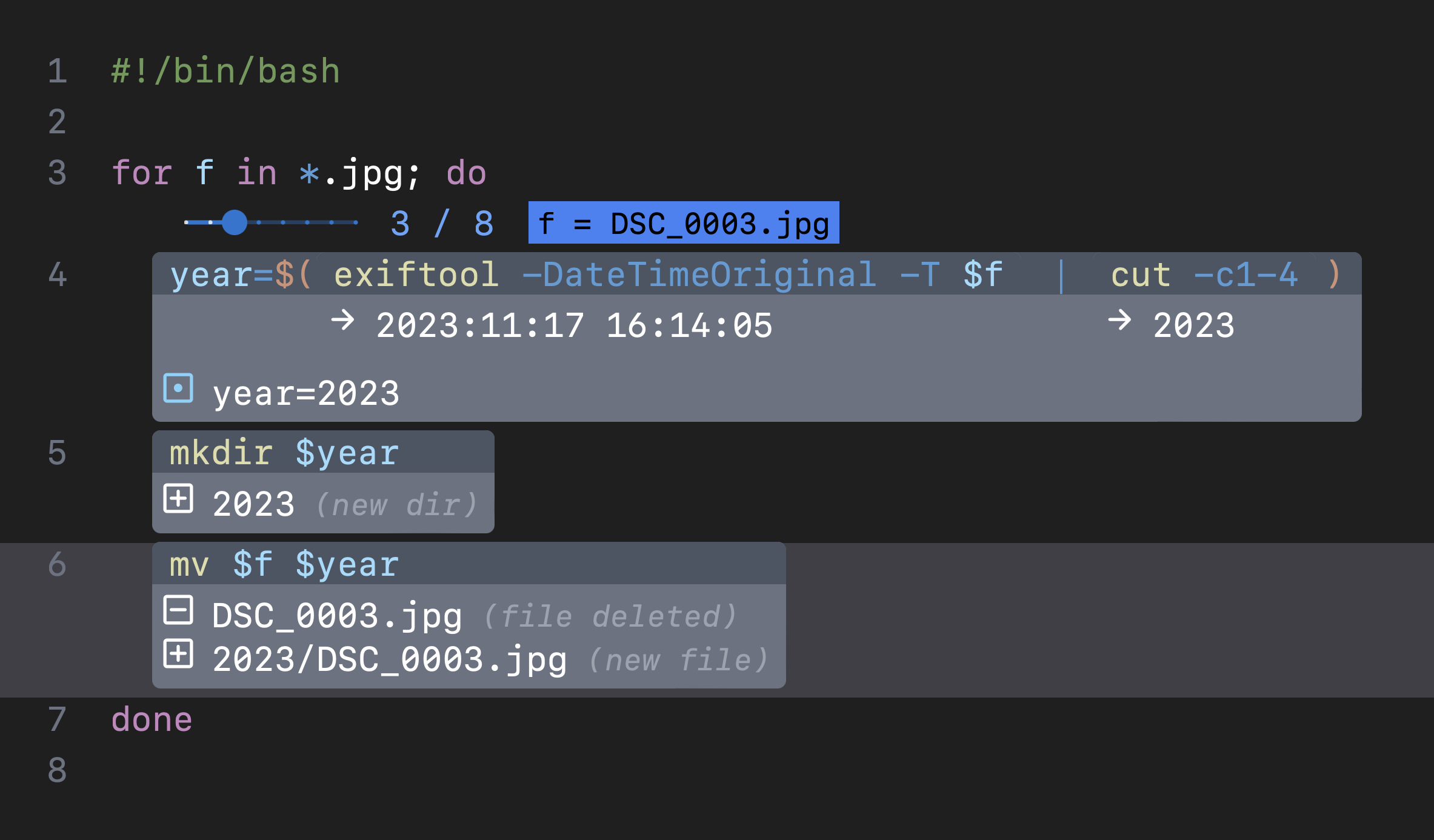}\caption{An earlier design for ShellVis which places bubbles ``inline'' with code, rather than off to the side.}\Description{A version of the ShellVis interface similar to those described earlier, but instead of having widgets and bubbles to the right, they are integrated into the flow of code. In particular, the bubble for a command surrounds the command itself; commands embedded in other commands (such as with "$()") create nested bubbles.}
\label{fig:inline}
\end{figure}

The ShellVis visualizer runs in a separate window from the code editor. Although this display is separate, it is essential that the programmer be able to shift their attention easily between their editor and the visualizer as they work. These shifts in attention become easier if the visualizer closely resembles code in the editor. (This is why, for instance, the visualizer matches Visual Studio Code's syntax highlighting.) But, in tension to this goal, we often have many lines of information for each line of code, such as a large output text or multiple file system modifications. Showing all this information pushes successive lines of code down, resulting in a mismatch between the editor and visualizer.
An earlier design for ShellVis, shown in Figure~\ref{fig:inline}, placed feedback bubbles ``inline'' with commands in code rather than off to the side, and suffered from this mismatch problem.

Projection Boxes \cite{doi:10.1145/3313831.3376494} faced a similar issue, and resolved it with per-line boxes that float to the right of the code. When the current line changes, the line's box moves to be aligned with it, while boxes above and below get pushed out of the way, often off the screen. A limitation of this approach is that it becomes difficult to survey the behavior of an entire program at a glance when most lines' boxes have been pushed entirely out of view. Put another way, Projection Boxes offer low peripheral visibility.

\begin{figure}
\centering
\includegraphics[width=0.90\linewidth]{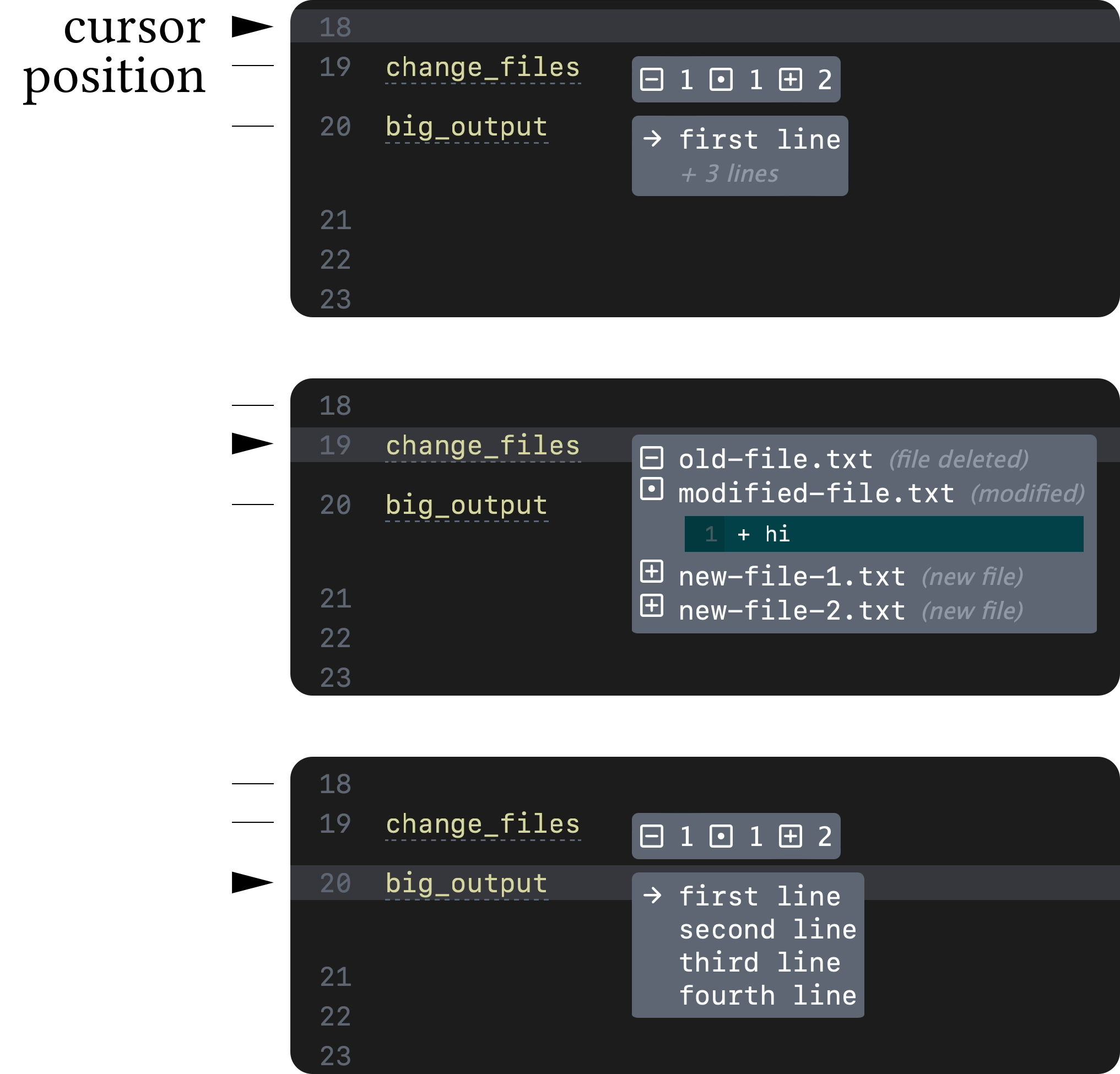}

\caption{Examples of how bubbles expand and collapse as the editor's cursor passes over them.}\Description{Three screenshots of the same code being viewed in ShellVis, but with different lines selected (which is shown with a highlight background). As a line is selected, the bubble to its right changes from an abbreviated form to an expanded form. A command "change_files" starts abbreviated in a one-line bubble that reports "minus 1, dot 1, plus 2", but expanding it shows a larger bubble that says specifically which files were deleted, modified, or added. A command "big_output" starts abbreviated with only the first line of output and the text "+ 3 lines", but expanding it shows its full output (4 lines).}
\label{fig:abbreviations}
\end{figure}

Our approach is instead to use ``abbreviated bubbles'' which expand to full bubbles when focused, a form of ``fisheye view'' \cite{doi:10.1145/22627.22342, doi:10.1145/1124772.1124921}.
As shown in Figure~\ref{fig:abbreviations}, abbreviated bubbles are one or two lines in height and show a summary of the full bubble's contents.
Bubbles expand as ``pop-ups'' that may cover up bubbles lower on the page, rather than move them. This helps to keep the overall visualization stable and predictable as focus changes.

As in Projection Boxes, visualization focus in ShellVis is generally controlled by moving the cursor in the editor. Since our visualizer is separate from the editor, a bit of machinery is needed to get cursor information from the editor to the visualizer. We made an extension for Visual Studio Code which reports this information to the ShellVis server via a WebSocket connection. Analogous extensions could be made for other editors. Bubbles can also be expanded in the visualizer by hovering over them with the mouse.

\subsection{Visualizing Loops}
\label{sec:visualizing-loops}

Another design challenge is to visualize the execution of loops. Loops create a structural mismatch between a program's source code and its trace: a single line of code in a loop may be executed many times. This presents a new instance of the problem above: how can this large amount of information be presented while maintaining a close correspondence with source code?

\begin{figure}
\centering
\includegraphics[width=0.90\linewidth]{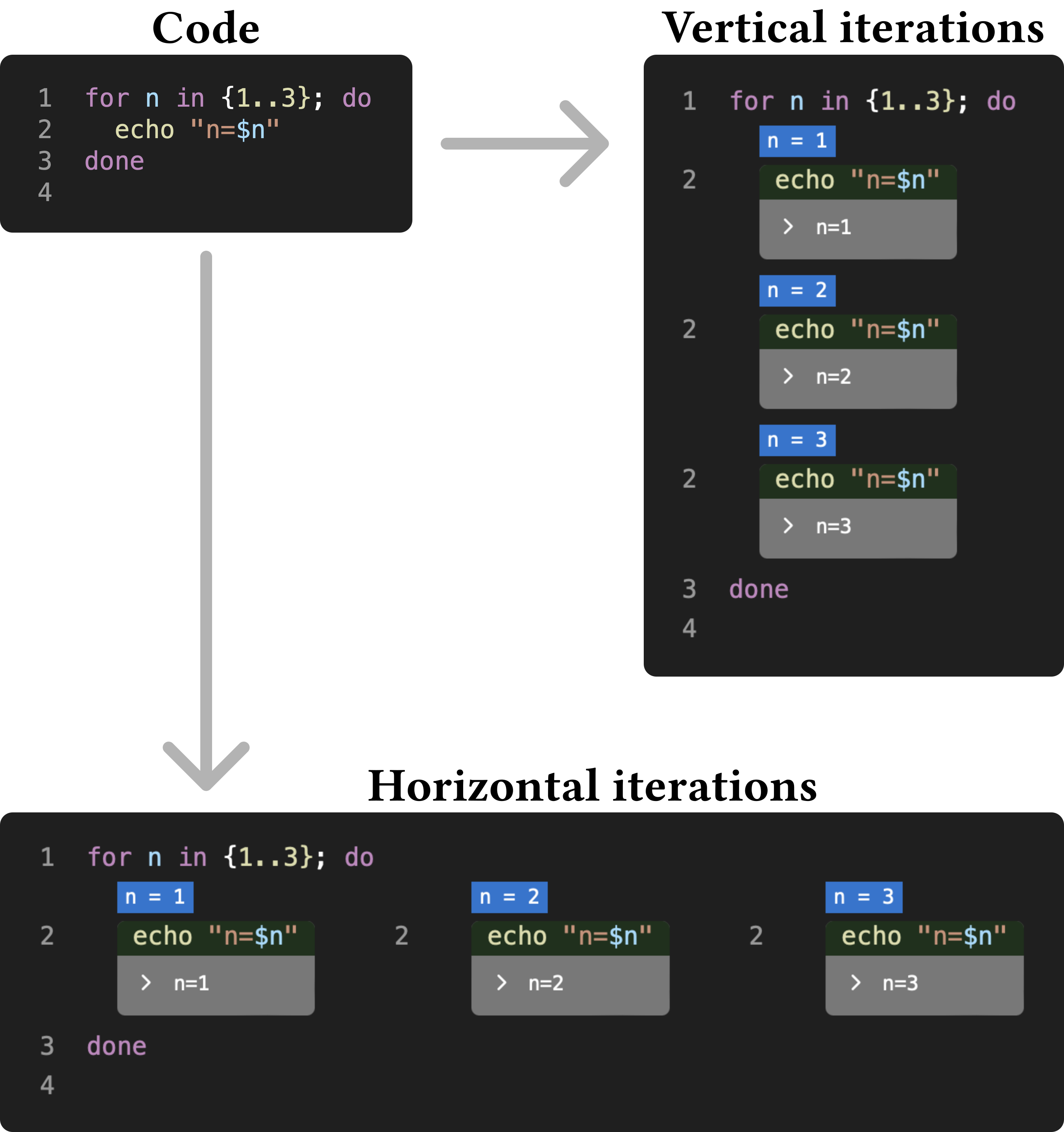}

\caption{Two alternative designs for loops explored during ShellVis's development which show all iterations simultaneously, rather than giving the user a slider to select one iteration at a time.}\Description{A screenshot labelled "code" shows a small script consisting of a 3-iteration loop with a single echo command inside it. A screenshot labelled "vertical iterations" shows a version of the inline-bubbles interface described above, but with three copies of the loop's body unfolded inside of it, showing three different bubbles for the echo command with three different outputs. A screenshot labelled "horizontal iterations" shows a similar idea, but with the three iterations laid out horizontally.}
\label{fig:loop-layouts}
\end{figure}

Early designs of ShellVis (shown in Figure~\ref{fig:loop-layouts}) showed ``expanded'' views of loops, laying out loop iterations either horizontally or vertically, repeating the loop code for each iteration, and using the ``inline'' version of command bubbles shown in Figure~\ref{fig:inline}. These designs appealed to us because they followed our general principle of showing as much information as possible by default, minimizing interaction and permitting comparisons across iterations with a mere movement of the eye.\footnote{As Victor~\cite{magic-ink} says, ``The hand is much slower than the eye''.} However, these designs disrupt the visual correspondence between the visualizer and source code in the editor, which is essential if the visualizer will be used in a tight loop with changes to code.

We ultimately opted to address this problem with interaction. As shown in Figure~\ref{fig:demo} and discussed in \S\ref{sec:shellvis-by-example}, ShellVis provides sliders for selecting iterations of interest inside \texttt{for} and \texttt{while} loops. As the user drags the slider, ShellVis shows either the value of the loop variable (for a \texttt{for} loop) or a command bubble for the loop condition (for a \texttt{while} loop). Annotations in the body of the loop update to reflect the selected iteration. Using the slider, the user can make sure the loop is behaving as expected across iterations.

\subsection{Interface Scalability}
\label{sec:interface-scalability}

The current UI design of ShellVis scales well to moderate amounts of data. If a loop has hundreds of iterations, the user can scrub through them as easily as if it had three. If a command has a large output, its output bubble will grow to a maximum size and then let the user scroll its contents. (One of our study tasks involves a file with several thousand lines.) We have not optimized our design or implementation for ``big data''; within a shell context, handling hundreds to thousands of iterations or lines is sufficient for many end-user applications.

\section{Implementation}
\label{sec:implementation}

ShellVis is implemented in two parts: a server and a client. The ShellVis server runs on the user's own system. It is responsible for running a shell script in a sandboxed environment, tracing its activities, and reporting this information to the client as a live-updating program trace. The client runs inside a web browser. It is responsible for displaying the program trace to the user and allowing the user to explore it interactively.

Implementation of the ShellVis client was fairly routine. As a shell script executes and is traced, the trace is streamed from the server to the client with Automerge \cite{automerge} and displayed with the help of React \cite{react}. The remainder of this section discusses the implementation of the ShellVis server.

\subsection{Sandboxing}
\label{sec:sandboxing}

To make live programming of shell scripts possible, ShellVis must sandbox and trace the effects of scripts. Currently, sandboxing is limited to effects on the filesystem. ShellVis currently does nothing to sandbox other effects, such as network requests. We discuss the prospects for these other effects in \S\ref{sec:sandboxing-more}.

ShellVis sandboxes the file system with a \emph{union file system} \cite{pendry1995union}. With a union file system, a user can mount a virtual copy of the root file system inside a temporary directory (the ``union directory'') such that reads go straight to the root file system but writes go to a ``delta'' directory which records file modifications. Subsequent reads of changed files route to the delta directory instead, making a union file system essentially a ``copy-on-write'' file system. If a shell script is run in the union directory rather than directly on the root file system, it can read the entire root file system, but any changes it performs will be recorded in the delta directory rather than touching real files. This protects the original file system.

The contents of the delta directory also tell us what file system modifications have occurred. As we wish to know more precisely what modifications are produced by \emph{each line} of a script, ShellVis actually stacks two union file systems on top of one another. Each command is run in the upper union file system. After it is run, the upper delta directory is scanned for changes to add to the program trace, and the changes are committed to the lower union file system. The upper delta directory is then cleared for the next command.\footnote{Although this approach generally works, it may fail to ascribe changes to the correct command if multiple commands write to the file system concurrently. This could be fixed by using a more sophisticated virtual file system, or by using tracing tools which directly target individual processes like dtrace \cite{dtrace} or fsatrace \cite{fsatrace}.}

As mentioned in section \S\ref{sec:shell-tools}, ShellVis's approach to sandboxing is directly inspired by \texttt{try}, a command-line tool for Linux which lets users sandbox individual commands \cite{try}. \texttt{try}'s sandbox is based on two Linux operating-system features: OverlayFS and namespaces. OverlayFS is a union file system built into Linux. Namespaces provide several important forms of process isolation. Most importantly for \texttt{try}, they make it possible to run a process with an arbitrary directory chosen as its root directory, so references to ``/'' in the program will resolve to that directory. \texttt{try} runs its target command in a namespace where ``/'' refers to the OverlayFS mount directory.

ShellVis is currently built for macOS, which does not have a built-in union file system. Instead, ShellVis uses unionfs-fuse \cite{unionfs-fuse}, which is built in user space on top of FUSE. macOS also does not have namespaces. It does provide the more primitive \texttt{chroot} call, which provides the root-setting function of namespaces without its other features. While it should be enough for our purposes, we ran into several issues with \texttt{chroot}. Using \texttt{chroot} on macOS requires disabling System Integrity Protection, which is cumbersome and weakens the computer's security. Even with SIP disabled, we were unable to make \texttt{chroot} work seamlessly on macOS. Without \texttt{chroot}, our sandboxing is leaky. File accesses performed with relative references are sandboxed, since they remain in the sandbox directory, but absolute paths will refer to original files outside the union file system. This disqualifies the current ShellVis prototype from being used in the wild by users who may be unaware of this limitation. Future work on ShellVis will likely move to Linux, where these problems can be more easily fixed and which is generally a more accommodating environment for OS-level experimentation.

\subsection{Tracing}
\label{sec:tracing}

The ShellVis server watches a target script file for changes and re-runs the script whenever the editor saves a new copy. This and the selection monitoring mentioned in \S\ref{sec:bubbles} are the only ways the editor communicates with ShellVis.

\begin{figure}[h!]
\centering
\nop{START-NO-COUNT}{\scriptsize \begin{minted}{sh}
echo "hi!"

↓

local sv_ret >/dev/null
local sv_uploadId_stdout sv_uploadId_stderr sv_uploadId_varsEnter\
  sv_uploadId_varsExit >/dev/null
sv_msg "{\"type\":\"call-enter\",\"nodeId\":\"CallExpr_1_1_1_11\",\
  \"context\":\"$(sv_ctx_str)\",\"cwd\":\"$PWD\",\"suppressed\":false}" |
  read -r sv_uploadId_stdout sv_uploadId_stderr sv_uploadId_varsEnter\
    sv_uploadId_varsExit
sv_typeset | sv_upload $sv_uploadId_varsEnter
echo "hi!" \
  1>&1 1> >(sv_upload $sv_uploadId_stdout) 2>&2 2> \
  >(sv_upload $sv_uploadId_stderr)
sv_ret=$?
sv_typeset | sv_upload $sv_uploadId_varsExit
sv_msg "{\"type\":\"call-exit\",\"nodeId\":\"CallExpr_1_1_1_11\",\
  \"context\":\"$(sv_ctx_str)\",\"cwd\":\"$PWD\",\"exitCode\":\"$sv_ret\"}" \
  >/dev/null
sv_exitcode $sv_ret
\end{minted}

}\caption{Example of how ShellVis transforms a minimal one-line shell script. The transformed script also includes a prelude defining the \texttt{sv\_} functions, not shown here.}\Description{A script containing the single line 'echo "hi!"' becomes many lines of instrumented code. (No image file.)}
\label{fig:transform}
\end{figure}

The server needs to somehow obtain detailed, line-by-line information about a shell script's behavior when it runs. Our approach here was to transform the script before running it to add additional tracing instructions. An example of this transform is shown in Figure~\ref{fig:transform}. The \texttt{sv\_msg} and \texttt{sv\_upload} functions are used in this transformed code to report events (like a command returning) and upload data (like standard IO streams from a command) to the ShellVis server over one-off HTTP connections. This is used for tracing, as well as to manage per-command sandboxing.

To perform the source transformation, ShellVis uses the mvdan/sh library \cite{mvdan-sh} to parse the script to an AST and then walks this AST to insert tracing code. If this parsing fails, the library often returns more useful error messages than the shell itself, which we report to the user. Currently, ShellVis instruments calls, for loops, and while loops. A future version of ShellVis could instrument more (conditionals, functions, etc.), but these were omitted in our prototype. A script with these features will run correctly in ShellVis, but with an incomplete trace.

An alternative approach to tracing a script's execution would be to fork a shell's codebase and add instrumentation directly to its interpreter, so an unmodified script can produce a trace when run. Through our implementation work, we have discovered that transforming a shell script without modifying its behavior can be a complex and error-prone endeavour, so instrumenting a shell directly may be preferable for future work.

\subsection{Performance}
\label{sec:performance}

ShellVis currently imposes a significant performance overhead over running a script directly. For instance, the final script constructed in section \S\ref{sec:shellvis-by-example} takes around 3.5 seconds to run in ShellVis, and only 0.5 seconds if run directly. Profiling suggests that most of this overhead comes from communication between the instrumented shell script and the ShellVis server. Earlier in development, we explored more performant approaches to script-server communication, including named pipes and persistent TCP connections, but HTTP proved to be more reliable at this prototype stage. Moving the ShellVis server inside the shell (as described at the end of \S\ref{sec:tracing}) would make lower-overhead communication easier to implement.

ShellVis relies on unionfs-fuse for sandboxing, so script commands that read or write to the file system must go through that system. We expect this incurs an additional cost. If ShellVis were moved to Linux, it would be able to use OverlayFS, which is likely faster since it runs in the kernel rather than in user space.

ShellVis's approach of re-running a script from scratch whenever it changes will become less practical for larger or slower scripts, not due to overhead from ShellVis but due to the intrinsic cost of running the script. Consider scripts that process large media files like videos, or those that search through large directory structures. New techniques could make it more practical to live-program slower scripts like these. For instance, loops could automatically be limited to a smaller number of iterations for testing, or the results of long-running subcommands could be cached.

\section{User Study Design}
\label{sec:experiment-discussion}

In order to assess whether sandboxed live-programming achieves its goals in the ShellVis prototype, and to provide further insight into the experiences of shell programmers, we conducted a user evaluation.
Our study focused on three research questions:

\begin{itemize}
\item \textbf{RQ1:} What challenges do programmers face working with shell scripts?
\item \textbf{RQ2:} How do programmers make use of live feedback when working with shell scripts?
\item \textbf{RQ3:} How do programmers make use of sandboxing when working with shell scripts?
\end{itemize}

\subsection{Participants}
\label{sec:participants}

We recruited seven participants through social media channels. Shell-scripting experience ranged from 1 to 20 years. All our participants were men between the ages of 28 and 42.

\subsection{Procedure}
\label{sec:procedure}

A study session began with a short interview focused on the participant's history with shell-scripting and problems they ran into while working with shell scripts. We deliberately opened the session with this interview in order to obtain a participant's prior perspective, unbiased by the approach of our prototype.

Next, we taught the participant how to use ShellVis through the photo-sorting scenario shown in \S\ref{sec:shellvis-by-example}.
We ran ShellVis on our computer, accessed by the participant through Zoom's remote control feature.
We then asked participants to perform four tasks:

\begin{itemize}
\item \textbf{titles:} Rename text files to the titles contained in their first lines, with a possible pitfall caused by the presence of spaces in titles. Starts from a blank file.
\item \textbf{users:} Delete user directories named in lines of a text file, with a possible pitfall caused by a blank line in the file. Starts from faulty code, such as might be produced by a code assistant.
\item \textbf{log:} Read a complex pipeline which filters a log and aggregates information from it.
\item \textbf{lines:} Multiply the line-counts of files in a folder.
\end{itemize}

\begin{figure}[t!]
\centering
\includegraphics[width=\linewidth]{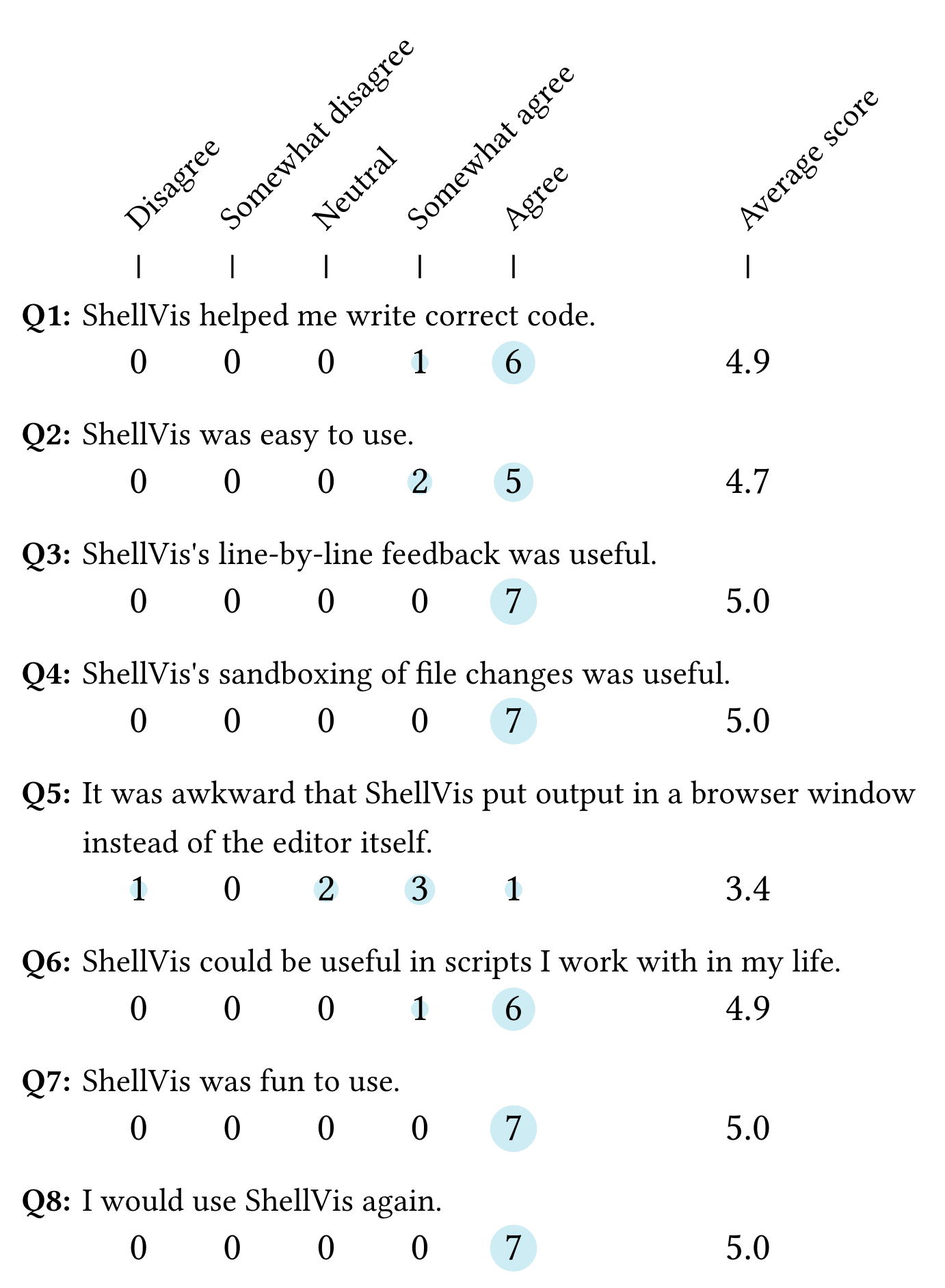}\caption{Questions in survey along with response distributions among the 7 participants. All questions are on a 5 point Likert scale with 1 being “Disagree'' and 5 being “Agree''. Average scores are shown to the left.}\Description{A table of survey response counts. Possible survey responses are "Disagree", "Somewhat disagree", "Neutral", "Somewhat agree", and "Agree". The table shows counts for each question & response; here we will only report average scores. Q1: ShellVis helped me write correct code. Average is 4.9. Q2: ShellVis was easy to use. Average is 4.7. Q3: ShellVis's line-by-line feedback was useful. Average is 5.0. Q4: ShellVis's sandboxing of file changes was useful. Average is 5.0. Q5: It was awkward that ShellVis put output in a browser window instead of the editor itself. Average is 3.4. Q6: ShellVis could be useful in scripts I work with in my life. Average is 4.9. Q7: ShellVis was fun to use. Average is 5.0. Q8: I would use ShellVis again. Average is 5.0.}
\label{fig:survey}
\end{figure}

\noindent{} Full prompts for these tasks are available in Appendix \ref{sec:study-tasks}. We encouraged participants to use Google or other resources naturalistically if needed, though we did not allow use of LLM assistants like Copilot and ChatGPT. (See \S\ref{sec:gulfs} for some reflections on code assistants.)
After completing these tasks, the participant filled out a survey, with questions and responses shown in Figure~\ref{fig:survey}.

Finally, we conducted a post-study interview. This interview was not formally structured, but we focused on eliciting the participant's impressions of ShellVis and how they thought it might or might not fit into their real-world workflows. In particular, we asked participants whether they thought ShellVis would address any challenges they mentioned at the start of the session, or not.
Sessions ranged from one to two hours in length.

\subsection{Limitations}
\label{sec:limitations}

Our study has significant limitations. The tasks we assigned to users were simple and designed to make use of ShellVis's features. Our sample size is small and demographically narrow. While no participants were prior acquaintances of the authors, recruitment through social networks likely biased the study population towards people sympathetic to our goals. Since all participants had a least a year of shell experience, conclusions should not be drawn from this study about the value of ShellVis to shell-scripting novices.

\section{User Study Results}
\label{sec:results}

To analyze the study data, the first author performed open coding on video recordings, akin to phases 1-2 of thematic analysis as described in Clarke \& Braun~\cite{doi:10.1080/17439760.2016.1262613}. We then sorted these under headings corresponding to our research questions and identified themes within each. The remainder of this section reports the results of this analysis. Quotes from participants are highlighted \sq{``like this''} and labelled with participant IDs (like \textbf{P}\textsubscript{7}).

\subsection{Challenges with Shell Scripts (RQ1)}
\label{sec:rq1}

When we asked participants in our initial interview about problems they had run into with shell scripts, most participants were quick to recount difficulties.
A few had pointedly negative feelings, responding that \sq{``everything''} (\textbf{P}\textsubscript{2}) or \sq{``essentially every part of {[}shell scripting{]}''} (\textbf{P}\textsubscript{3}) was challenging.

The most common problems mentioned were not knowing, or not remembering, how to do certain things in shell scripts.
\sq{``Can you remember what the right thing to call is? Because there's so many magic spells\ldots{} {[}even{]} if you know the spell, how do you cast it?''} (\textbf{P}\textsubscript{1}).
\sq{``I use {[}shell scripts{]} only frequently enough to remember that there is a way to do something, but I don't remember exactly what is the way.''} (\textbf{P}\textsubscript{3}). \sq{``Every time, I reach for Google first, or now ChatGPT.''} (\textbf{P}\textsubscript{5}).
\sq{``{[}I{]}t's a matter of, how do you bring up the right incantation to get it working right\ldots{} I end up reaching for the manual {[}more{]} often than not\ldots{}''} (\textbf{P}\textsubscript{6}).
\sq{``I find some of the shell keywords quite challenging to remember\ldots{} these things I constantly have to look up even though I write these time and time again.''} (\textbf{P}\textsubscript{7}).
Participants often attributed these memory problems to not using shell scripts often. Shell scripts seem to occupy an unfortunate middle ground: used too infrequently to be remembered, but frequently enough that forgetting is a pain.
Participants employ different techniques to get around these problems, including Google, ChatGPT, looking up examples of one's prior work, and man pages. These ``gulf-of-execution'' struggles have a subtle relationship to ShellVis's ``gulf-of-evaluation'' approach \cite{doi:10.1201/b15703}. We discuss this issue further in \S\ref{sec:gulfs}.

Several participants reported they had run into disasters while working with the shell: \sq{``I'm extremely paranoid\ldots{} because I have seen files disappear. And there's nothing you can do about it later; it's pretty hard to recover.''} (\textbf{P}\textsubscript{3}). To protect themselves against such disasters, shell-scripters employ different methods. \textbf{P}\textsubscript{3}, \textbf{P}\textsubscript{4}, and \textbf{P}\textsubscript{7} mentioned putting \texttt{echo} in front of potentially destructive commands to ``dry run'' them. Other participants tried to backup files before starting work on a script. \textbf{P}\textsubscript{5} says \sq{``always make backups''}, but also reported that backups are annoying to use, take up space, and you may simply forget to make them.

In the initial interview, participants did not generally mention lack of visibility, the problem live programming most directly addresses, as a problem. The one exception was \textbf{P}\textsubscript{7}, who remarked:
\sq{``As a shell script gets more and more complicated, there's no observability into these pipelines\ldots{} I think the lack of observability\ldots{} is sometimes a little bit of a friction point.''}
Problems with lack of visibility were mentioned more often as participants worked through tasks with ShellVis and reflected afterwards.

\subsection{Using Live Feedback (RQ2)}
\label{sec:rq2}

Participants unanimously agreed that ShellVis's line-by-line feedback was helpful (Q3 avg of 5.0/5). As ShellVis is at its heart a visualizer, it is perhaps not surprising that participants found ShellVis useful for understanding code they hadn't written (as specifically probed in \textbf{logs} and to a lesser extent \textbf{users}):
\sq{``I think I kind of know what that regex does, but I'm not sure\ldots{} {[}ShellVis{]} is nice because it gives you the output right there\ldots{}''} (\textbf{P}\textsubscript{1}),
\sq{``If I were looking into a shell script that I didn't author myself, {[}ShellVis{]} would also be incredibly valuable\ldots{} {[}Without ShellVis,{]} you only see the pieces of the pipeline, but not the data flowing through.''} (\textbf{P}\textsubscript{7}).

Participants also found ShellVis useful while actively writing code, to plan their next step or diagnose why their last step failed. \textbf{P}\textsubscript{7} explained this experience: \sq{
``{[}With ShellVis,{]} when I'm reasoning about a future step, I can look at intermediate artifacts from previous steps without having to rerun. I could scan upward on the web view and be like: oh yeah, that variable I used earlier\ldots{} it has this data, versus having to reload that context in my head after I had gone maybe two, three steps forward.''}. A specific example of this experience: While working on \textbf{titles}, \textbf{P}\textsubscript{4} wondered why a use of \texttt{\$file} in their code wasn't producing the expected result.
They asked what value \texttt{\$file} was taking on: \sq{``Can I see what \texttt{\$file} is for each\ldots{} Oh yeah, duh, it's right there!''}

Several participants described how live feedback provided confidence or comfort: \sq{``The preview {[}shows{]} that when you're running this, the first file that gets picked up is this. That builds a certain confidence level within me that, oh, that's right.''} (\textbf{P}\textsubscript{6}), \sq{``Just like knowing that things are successfully running is very comforting.''} (\textbf{P}\textsubscript{7}). We will discuss these affective dimensions more in \S\ref{sec:rq3}.

\subsubsection{ShellVis vs. \texttt{echo} Logging}
\label{sec:shellvis-vs.-echo-logging}

A natural point of comparison for ShellVis is logging information from a script with \texttt{echo}. Participants agreed that ShellVis's output annotations were a significant improvement over this status quo, and gave several reasons why.

First, automatic output annotations remove the tedium of adding and removing logs during a debugging process: \sq{``I no longer have to litter my {[}script{]} with echoes that I need to subsequently search for and clean up.''} (\textbf{P}\textsubscript{7}). In response to a question about \texttt{echo}-based logging, \textbf{P}\textsubscript{4} reported: \sq{``You can put all this crap in\ldots{} and have it print all of it, but then you have to put in case handling so that it isn't printing all the time, because you don't want all this garbage debugging stuff going across your console.''}.

\textbf{P}\textsubscript{5} further noted that ShellVis's ``show everything'' approach obviates the need to know \emph{where} to place logs: \sq{``I think it's huge because you're removing a whole extra step\ldots{} knowing where to put the right echoes. I don't have to think about that because it's all there. {[}With logs{]} I wouldn't really know which line is causing the problem. It could be a line further up that's causing a downstream problem, and I could be putting the echoes in the wrong place.''}

Finally, participants reported numerous ways it was more helpful to see output annotations in context, attached to specific lines of code and structures like loop iterations, than to see echoed output streaming in a terminal. \textbf{P}\textsubscript{3} reported that, with \texttt{echo}-based logging, \sq{``finding exactly where you are is difficult\ldots{} if your logs are too sparse, mapping that back to the source can be difficult\ldots{} Or if it's in a for loop, at what execution did that happen?''}. \textbf{P}\textsubscript{4} similarly reported that \sq{``It's easy when you're a couple of nested loops in to get lost on where you are. So to see {[}with ShellVis{]} what the machine thinks is very helpful.''}

\subsubsection{ShellVis vs. Debuggers}
\label{sec:shellvis-vs.-debuggers}

Debuggers did not arise in conversation as much as \texttt{echo} logging did, likely because debuggers are used less often by participants. \textbf{P}\textsubscript{4} had some experience with the PowerShell debugger, but explained that ShellVis was \sq{``a little bit {[}different{]}, cause to see what variables are {[}with a debugger{]}, you have to single-step stuff\ldots{} Whereas this is like, let's run it. And now I can just use the little slider thing to see as it was going along\ldots{} It's a good way to visualize what the shell script is doing.''}

\subsection{Using Sandboxing (RQ3)}
\label{sec:rq3}

Participants unanimously agreed that ShellVis's sandboxing of changes to the file system was helpful (Q4 avg of 5.0/5). In our initial interviews (\S\ref{sec:rq1}), participants described techniques they used in the past to protect themself from accidents with the shell. After trying ShellVis, several explained how ShellVis removed the tedium and error-proneness of implementing those workaround techniques. \textbf{P}\textsubscript{4} described how they typically put lots of work into into making ``dry run'' versions of scripts, and compared this to ShellVis: \sq{``If {[}ShellVis{]} was able to catch that kind of stuff as a sandboxing test and make the automated execution stop, that's a pretty powerful aid -- and something I usually have to manually add.''} \textbf{P}\textsubscript{7} described their approach: \sq{``What I would ordinarily do is\ldots{} pick one file to start and then set that as \texttt{\$f}, so I might do something \texttt{f=/docs/doc1.txt}. And then\ldots{} if one file does change destructively, I could easily roll back. Or alternatively, I might start off echoing everything and to be like, okay, that looks sensible. But with {[}ShellVis{]}, I can just very confidently run the entire script, which is very cool.''}

We were gratified to hear some participants using affective language like ``confidence'' and ``comfort'' when describing the experience of using ShellVis. \textbf{P}\textsubscript{6} said \sq{``{[}ShellVis{]} gives you a lot of confidence that, look, I can fix it before I actually run it. So I think the cognitive load\ldots{} goes away. I'll be less worried about executing this and making a mess of my system.''}. \textbf{P}\textsubscript{3} said \sq{``Having this safe environment makes me very comfortable with just wildly trying things and see what happens, versus I'm usually very, very cautious because deletions are very hard to undo.''}. This language of exploration and experimentation was echoed by \textbf{P}\textsubscript{7}: \sq{``{[}ShellVis{]} is an environment that allows this kind of experimenting. You can kind of be aggressive.''}

Affective dimensions like these may be an important (but difficult-to-measure) benefit of safe live-programming environments. Beyond making programming faster or less error-prone, these environments may cultivate the kind of play and experimentation that support learning and creativity. Our study participants universally reported that they had fun working with ShellVis (Q7 avg of 5.0/5).

Of course, a feeling of safety should be backed up by actual safety. Several participants made clear that ShellVis would need to earn their trust before they would feel comfortable using it (\textbf{P}\textsubscript{1}, \textbf{P}\textsubscript{4}).
As discussed in \S\ref{sec:sandboxing}, this wariness is warranted, as ShellVis's file system sandboxing is currently incomplete.

\subsection{Overall Impressions}
\label{sec:overall-impressions}

After using ShellVis, participants were uniformly positive about its helpfulness in accomplishing study tasks (Q1 avg of 4.9/5) and its potential usefulness in their own work (Q6 avg of 4.9/5). All participants expressed an interest in using ShellVis in their own lives: \sq{``I think it's really cool; I would love to have a tool like that''} (\textbf{P}\textsubscript{1}), \sq{``This is good enough that I actually would like to be able to run it on my computer''} (\textbf{P}\textsubscript{2}), \sq{``It already works amazingly well\ldots{} It felt, like, transparent,''} (\textbf{P}\textsubscript{3}), \sq{``I would definitely use a tool like this''} (\textbf{P}\textsubscript{4}), \sq{``Is this coming out anytime soon?''} (\textbf{P}\textsubscript{5}), \sq{``Do you have an ETA for when this tool might potentially be made available?''} (\textbf{P}\textsubscript{6}), \sq{``I really want this tool and I would use it immediately.''} (\textbf{P}\textsubscript{7}).

\section{Discussion \& Future Work}
\label{sec:discussion-future-work}

\subsection{Gulf of Evaluation vs. Gulf of Execution}
\label{sec:gulfs}

As discussed in \S\ref{sec:rq1}, the most common problems participants reported from their shell-scripting experience had to do with not knowing how to express their intent as shell code, whether because they had never learned the shell's syntax or because they forgot it. In the language of Norman \& Draper~\cite{doi:10.1201/b15703}, these problems come from the \emph{gulf of execution}. At least at first glance, this seems quite unrelated to ShellVis's approach. ShellVis is designed to help programmers understand what a shell script is doing, a clear instance of bridging the \emph{gulf of evaluation} \cite{doi:10.1201/b15703}. In other words: Once you type something, ShellVis can help you see what it does, but it can't help you figure out what to type in the first place.

We asked participants about this in our final interview. They tended to agree with the interpretation above: \sq{``My initial response\ldots{} is that I don't think it helps there. Because to get the tool to do anything, you have to give it some input.''} (\textbf{P}\textsubscript{1}), \sq{``If I start off with the fact that I don't know how to structure a \texttt{for} loop, then I don't think {[}ShellVis{]} provides enough feedback to say, okay, this way.''} (\textbf{P}\textsubscript{6}). \textbf{P}\textsubscript{5} agreed, but added some nuance, suggesting that bridging the gulf of evaluation could in some situations assist in bridging the gap of execution: \sq{``Partially, maybe {[}ShellVis{]} helps because you see the error messages right there, and you can glean some information from the error messages\ldots{} but it's always better to know the commands beforehand.''}

In a broader sense, it is clear from participant reports in \S\ref{sec:rq2} and \S\ref{sec:rq3} that ShellVis \emph{did} help participants with the larger ``execution'' task of writing shell scripts, whether or not it was able to assist with low-level aspects like syntax and command flags.

LLM-powered code assistants, like Copilot \cite{copilot} and ChatGPT \cite{chatgpt}, are an emerging class of systems intended to bridge the gap of execution. While we did not test the use of ShellVis alongside code assistants, several participants expressed excitement that ShellVis could be a useful adjunct, especially with its sandboxing for extra protection. \sq{``I do think a common mode with {[}code assistants{]}\ldots{} is like, Oh, great. I wrote this code. It looks right. Is it right? I don't know. I guess I better check. And in that sense, I do think the the introspection afforded with this tool is pretty useful.''} (\textbf{P}\textsubscript{2})
\sq{``If {[}the code assistant{]} is wrong\ldots{} with your tool I'll see it immediately without destroying anything\ldots{} So I would certainly use LLMs more if I'm using ShellVis, because mistakes won't be costly.''} (\textbf{P}\textsubscript{3})
\sq{``Because I don't trust code assistants, there's still that verification phase\ldots{} Because this tool provides me live feedback, it's a very, very good way of verifying what the code assistant has put out. And the fact that it does it in a sandbox manner without a destructive mode is the key highlight.''} (\textbf{P}\textsubscript{6})
Further study is required to determine how well ShellVis supports the use of LLM-powered code assistants. We are encouraged by the results of Ferdowsi et al.~\cite{doi:10.1145/3613904.3642495}, who find positive results testing the use of LLMs alongside Projection Boxes. Shell scripting is riskier than the contexts those researchers studied, so the need to validate LLM-generated code with (safe) live programming may be even greater.

\subsection{System improvements}
\label{sec:improvements}

Interviews with participants surfaced numerous opportunities to improve ShellVis's design and feature-set.

\textbf{Editor integration.}
We were worried that our decision to run the ShellVis visualizer in a window separate from the editor (motivated in \S\ref{sec:scope}) would produce an awkward experience for users. Survey results partially support these concerns (Q5 avg of 3.4/5). \textbf{P}\textsubscript{2} and \textbf{P}\textsubscript{5} mentioned that the double-window arrangement took up valuable screen space they'd ordinarily use for other purposes, and \textbf{P}\textsubscript{2} was sometimes thrown off by the visual similarity of the two windows and tried to edit through the visualizer pane. Otherwise, this issue did not arise much in discussion, though that may be a product of the laboratory setting. A direction to explore in future work would be integrating ShellVis's line-by-line annotations into an existing text editor so that a second window is unnecessary. As discussed in \S\ref{sec:scope}, we deliberately chose not to do this for our first prototype, but ShellVis's client-server architecture should make further experiments with editor integrations straightforward.

\textbf{Surveying across loop iterations.}
Several participants noted that, though the loop scrubbing interaction was easy to use, it meant that anomalous behavior could hide in iterations that weren't visible at the moment. \sq{``Even though the for-loop scrubbing was cool, it would be nice to be able to see each loop in aggregate\ldots{} If I could just quickly scan through a list and see all loop iterations in some kind of way that grouped each iteration, I think that {[}spotting an anomaly{]} would have also been really easy\ldots{}''} (\textbf{P}\textsubscript{7}). Different approaches may accommodate the need to \emph{survey} many iterations of a loop at once. For instance, ``unfolded'' designs like those shown in Figure~\ref{fig:loop-layouts} or grid views like those used in Projection Boxes \cite{doi:10.1145/3313831.3376494} could show annotations for all iterations simultaneously. Alternatively, the existing slider could be maintained, but sparklines like those used in Hoffswell et al.~\cite{doi:10.1145/3173574.3174106} could be added to provide just enough ``peripheral vision'' for anomalies to be spotted from afar.

\textbf{Connecting lines across commands.}
While the loop view in ShellVis makes it easy to see chains of cause and effect within an iteration, idiomatic shell usage often involves commands that process streams line-by-line without loops, like \texttt{sed} and \texttt{awk}. (Our \textbf{log} task highlighted commands like this.) When line-by-line commands are piped together, ShellVis shows the outputs of each stage of the pipeline in separate bubbles, distancing corresponding input \& output lines. \textbf{P}\textsubscript{3} and \textbf{P}\textsubscript{7} both noted that this made it hard to follow the transformation of lines by commands, and they wanted a way to \sq{``go from a line output of a command line program and trace it back to an earlier command's line''} (\textbf{P}\textsubscript{7}). This desire makes perfect sense from a user's perspective, but it may be challenging to implement because Unix commands do not expose information on mappings between their input and output lines, and these relationships vary between tools.

\textbf{Extensible visualizers and editors.}
While ShellVis's generic annotation bubbles are usually enough to show what a shell script is doing, domain-specific visualizations for particular commands or situations could go even further. For instance, \textbf{P}\textsubscript{5} often writes scripts that process JSON data which they need to inspect. They currently use Postman \cite{postman} for this, which provides a better interface than the shell for looking at data, but which requires moving away from their script into another interface. As another example, \textbf{P}\textsubscript{7} mentioned wanting to color-code log output to more easily spot patterns in a time series, and to visualize 3D point clouds processed by another script. These examples suggest the potential of a system whereby ShellVis's visualizations could be extended for specific commands or output types. As a lighter-weight alternative to a de novo extension system, \textbf{P}\textsubscript{7} proposed that shell scripts could make their own visualizations by generating HTML which ShellVis would display in rendered form. Beyond visualizing output, extensions to ShellVis could change how parts of a script are \emph{authored}. \textbf{P}\textsubscript{1} suggested \sq{``a GUI for any command line -- as long as there's a tool that someone likes, or if you're a developer of a tool, you could write some sort of GUI spec that plugs into ShellVis.''}. \textbf{P}\textsubscript{7} imagined a system where the programmer could directly edit a command's output and the system could infer a change to the command to match the desired manipulation---a vision similar to output-directed programming \cite{doi:10.1145/3332165.3347925}. Systems like livelits \cite{doi:10.1145/3453483.3454059} and Engraft \cite{doi:10.1145/3586183.3606733} embed live GUIs inside of code, suggesting pragmatic ways to achieve these ``richer'' visions \cite{lrc}.

\textbf{More shell features.}
ShellVis does not yet provide feedback for the full gamut of shell features, as mentioned in \S\ref{sec:tracing}. The tasks we asked participants to perform did not call for the features ShellVis does not support, so these limitations did not arise much in our interviews. The one limitation noted by a study participant was when \textbf{P}\textsubscript{2} was working on \textbf{titles}: \sq{``It would be cool to see not only the command I'm running {[}in the source file{]}, but also what actually gets run in the post interpolation step.''}. We agree, as this feature would help illuminate many mistakes from misuse of the shell's idiosyncratic interpolation and quotation features.

\textbf{Remote use.}
At least three participants (\textbf{P}\textsubscript{2}, \textbf{P}\textsubscript{3}, \textbf{P}\textsubscript{4}) wondered if ShellVis could be used to edit scripts running on remote systems. While ShellVis does not currently support this, its client-server architecture should make it easy to adapt to this use: the server can run on a remote system while the client runs locally in a user's browser.

\subsection{Sandboxing More Effects}
\label{sec:sandboxing-more}

Not all side effects of shell scripts take place through the file system. The second-most prominent side effect of scripts is probably network requests. A network request can be patently side-effectful, such as a request to a database server to delete a user's account. Even seemingly innocuous requests, such as to HTTP GET endpoints, might cause problems if made indiscriminately, like costing money for a paid API or triggering rate limits. Beyond the file system and network requests, there are many other ways scripts can cause effects: writing to the clipboard, modifying running processes, making system calls, or causing physical effects like a print job.

These effects call for different approaches to sandboxing. We might call our current file system approach \emph{simulation}, as we are able to, with high fidelity, simulate a script's effects on the file system using a union file system. In contrast, it is impossible to simulate network requests to arbitrary hosts, as we cannot anticipate, in general, how remote hosts might respond to requests.

When simulation is not possible, the simplest alternative approach is \emph{detecting and quarantining} side effects. This involves detecting that a command will have an effect, stopping execution, showing the user the intended effect (in the context of the program trace), and giving them options of how to proceed. If the effect is entirely unexpected and unwanted, they can stop execution and revise the program to fix the problem. If they deem the effect harmless, they can approve it and execution will continue. If the effect looks like the right effect to perform, but the user doesn't want to perform it yet due to irreversible side-effects (say, on a remote host), they can mock it: provide, by hand, what they anticipate the response will be so that execution can continue. On top of the manual interactions described above, layers of automation are possible: commands could be placed on allowlists and denylists, or mock utilities could be provided to automate the mocking of a command.

We have not tested any of these approaches yet. Encouragingly, \textbf{P}\textsubscript{4} (a system administrator who frequently writes \sq{``potentially destructive''} scripts) brought up some of these ideas themselves in our conversation: \sq{``If there was a way to have the tool say, okay, I would have run this, we're going to assume that it worked and set the variable to this predefined thing so that the script execution can continue.''} \textbf{P}\textsubscript{4} was the only participant who brought up the need to sandbox more than the file system. This should not be taken as a sign that the issue is unimportant; we expect our study design did not prime participants to think about this issue.

\subsection{Sandboxed Live Programming Beyond Shell Scripts}
\label{sec:sandboxed-live-programming-beyond-shell-scripts}

We believe the scope of the \emph{sandboxed live programming} technique is much broader than just shell scripts, as practical situations typically call for side-effects. The need for sandboxing may be especially strong in end-user programming contexts, both because end users' concerns are often tightly tied to immediate effects they wish to perform, and because end users often have less experience programming and require more support.

How might the sandboxing approach of ShellVis translate from the Unix world of files and command-line programs to the GUI world? For instance, AppleScript \cite{doi:10.1145/1238844.1238845} and Shortcuts \cite{shortcuts} are programming systems for automating graphical applications. What would it take to sandbox these application-centered programming systems? Could it enable a higher degree of live feedback, as it did for ShellVis?

Beyond familiar computing environments, new possibilities on the horizon may enable richer forms of sandboxing. Local-first software \cite{doi:10.1145/3359591.3359737} reproduces the conveniences of modern cloud computing while allowing transparent local access to data. CRDTs, the data structure underlying many local-first systems, innately support branching and merging, capabilities which could be used to sandbox scripts. Local-first software may be a fertile substrate for broader use of sandboxed live programming.

\subsection{An Environment for Learning}
\label{sec:learning}

Our study participants were proficient programmers. We don't expect that someone without shell-scripting experience would perform well (or have a good time) in our study, especially given how little ShellVis does to help bridge the gap of execution (\S\ref{sec:gulfs}).
Nevertheless, we hope that ShellVis might make shell scripting more accessible to novices, as an adjunct to other learning resources. Live-programming environments have previously shown promise in learning contexts \cite{doi:10.1145/3478431.3499305, doi:10.1145/3126594.3126632}. We speculate that, by removing the risk of an irreversible mistake, ShellVis's sandboxing could make it easier for novices to learn through experimentation. \textbf{P}\textsubscript{4} joined our speculations: \sq{``{[}ShellVis{]} would make scripting a lot more accessible to people, I think, because you can still build iteratively, but now it's easier to see what each iteration would be doing.''}

\section{Conclusion}
\label{sec:conclusion}

The reach of live programming has been limited by its incompatibility with side effects. We have presented a general approach to this problem, \emph{sandboxed live programming}, and tested the approach through \emph{ShellVis}, our prototype of sandboxed live shell scripting. ShellVis combines overlay file systems with focus+context visualizations to create a new way of interacting with the shell, with the immediacy of the command line, the rich feedback of a notebook, and the program-editing capabilities of an editor. Participants in our study found the live visibility enabled by sandboxing helpful in understanding and writing code, and they enjoyed the feeling of safety provided by sandboxing. Discussion with participants also revealed the importance of complementing ShellVis with other tools to completely bridge the gulf of execution. With the development of further forms of sandboxing, these learnings can be extended to help make live programming practical in diverse end-user contexts with side effects.


\bibliographystyle{ACM-Reference-Format}
\bibliography{ms}


\begin{thebibliography}{59}


\ifx \showCODEN    \undefined \def \showCODEN     #1{\unskip}     \fi
\ifx \showDOI      \undefined \def \showDOI       #1{#1}\fi
\ifx \showISBNx    \undefined \def \showISBNx     #1{\unskip}     \fi
\ifx \showISBNxiii \undefined \def \showISBNxiii  #1{\unskip}     \fi
\ifx \showISSN     \undefined \def \showISSN      #1{\unskip}     \fi
\ifx \showLCCN     \undefined \def \showLCCN      #1{\unskip}     \fi
\ifx \shownote     \undefined \def \shownote      #1{#1}          \fi
\ifx \showarticletitle \undefined \def \showarticletitle #1{#1}   \fi
\ifx \showURL      \undefined \def \showURL       {\relax}        \fi
\providecommand\bibfield[2]{#2}
\providecommand\bibinfo[2]{#2}
\providecommand\natexlab[1]{#1}
\providecommand\showeprint[2][]{arXiv:#2}

\bibitem[{Apple Inc.}(2024)]%
        {shortcuts}
\bibfield{author}{\bibinfo{person}{{Apple Inc.}}}
  \bibinfo{year}{2024}\natexlab{}.
\newblock \bibinfo{title}{Shortcuts {User} {Guide} - {Apple} {Support}}.
\newblock
  \bibinfo{howpublished}{https://support.apple.com/guide/shortcuts/welcome/ios}.
\newblock


\bibitem[{Automerge contributors}(2024)]%
        {automerge}
\bibfield{author}{\bibinfo{person}{{Automerge contributors}}.}
  \bibinfo{year}{2024}\natexlab{}.
\newblock \bibinfo{title}{Automerge {CRDT}}.
\newblock \bibinfo{howpublished}{https://automerge.org/}.
\newblock


\bibitem[Bernstein(2024)]%
        {bashdb}
\bibfield{author}{\bibinfo{person}{Rocky Bernstein}.}
  \bibinfo{year}{2024}\natexlab{}.
\newblock \bibinfo{title}{Pipecut}.
\newblock \bibinfo{howpublished}{https://bashdb.sourceforge.net/}.
\newblock


\bibitem[Cantrill et~al\mbox{.}(2004)]%
        {dtrace}
\bibfield{author}{\bibinfo{person}{Bryan~M Cantrill},
  \bibinfo{person}{Michael~W Shapiro}, {and} \bibinfo{person}{Adam~H
  Leventhal}.} \bibinfo{year}{2004}\natexlab{}.
\newblock \showarticletitle{Dynamic {Instrumentation} of {Production}
  {Systems}}. In \bibinfo{booktitle}{\emph{Proceedings of the {General}
  {Track}: 2004 {Usenix} {Annual} {Technical} {Conference}}}.
\newblock


\bibitem[Chu(2024)]%
        {ysh}
\bibfield{author}{\bibinfo{person}{Andy Chu}.} \bibinfo{year}{2024}\natexlab{}.
\newblock \bibinfo{title}{Oils}.
\newblock \bibinfo{howpublished}{https://www.oilshell.org/}.
\newblock


\bibitem[Clarke and Braun(2016)]%
        {doi:10.1080/17439760.2016.1262613}
\bibfield{author}{\bibinfo{person}{Victoria Clarke} {and}
  \bibinfo{person}{Virginia Braun}.} \bibinfo{year}{2016}\natexlab{}.
\newblock \showarticletitle{Thematic analysis}.
\newblock \bibinfo{journal}{\emph{The Journal of Positive Psychology}}
  \bibinfo{volume}{12}, \bibinfo{number}{3} (\bibinfo{date}{dec 9}
  \bibinfo{year}{2016}), \bibinfo{pages}{297--298}.
\newblock


\bibitem[Cook(2007)]%
        {doi:10.1145/1238844.1238845}
\bibfield{author}{\bibinfo{person}{William~R. Cook}.}
  \bibinfo{year}{2007}\natexlab{}.
\newblock \showarticletitle{AppleScript}. In
  \bibinfo{booktitle}{\emph{Proceedings of the third {ACM} {SIGPLAN} conference
  on {History} of programming languages}}. ACM.
\newblock


\bibitem[Czapli{\' n}ski(2021)]%
        {up}
\bibfield{author}{\bibinfo{person}{Mateusz Czapli{\' n}ski}.}
  \bibinfo{year}{2021}\natexlab{}.
\newblock \bibinfo{title}{up - the {Ultimate} {Plumber}}.
\newblock \bibinfo{howpublished}{https://github.com/akavel/up}.
\newblock


\bibitem[D'Antoni et~al\mbox{.}(2017)]%
        {doi:10.1145/3106237.3106241}
\bibfield{author}{\bibinfo{person}{Loris D'Antoni}, \bibinfo{person}{Rishabh
  Singh}, {and} \bibinfo{person}{Michael Vaughn}.}
  \bibinfo{year}{2017}\natexlab{}.
\newblock \showarticletitle{NoFAQ: synthesizing command repairs from examples}.
  In \bibinfo{booktitle}{\emph{Proceedings of the 2017 11th {Joint} {Meeting}
  on {Foundations} of {Software} {Engineering}}}. ACM,
  \bibinfo{pages}{582--592}.
\newblock


\bibitem[Ferdowsi et~al\mbox{.}(2024)]%
        {doi:10.1145/3613904.3642495}
\bibfield{author}{\bibinfo{person}{Kasra Ferdowsi},
  \bibinfo{person}{Ruanqianqian~(Lisa) Huang}, \bibinfo{person}{Michael~B.
  James}, \bibinfo{person}{Nadia Polikarpova}, {and} \bibinfo{person}{Sorin
  Lerner}.} \bibinfo{year}{2024}\natexlab{}.
\newblock \showarticletitle{Validating {AI}-{Generated} {Code} with {Live}
  {Programming}}. In \bibinfo{booktitle}{\emph{Proceedings of the {CHI}
  {Conference} on {Human} {Factors} in {Computing} {Systems}}}. ACM,
  \bibinfo{pages}{1--8}.
\newblock


\bibitem[{fish-shell contributors}(2024)]%
        {fish}
\bibfield{author}{\bibinfo{person}{{fish-shell contributors}}.}
  \bibinfo{year}{2024}\natexlab{}.
\newblock \bibinfo{title}{fish shell}.
\newblock \bibinfo{howpublished}{https://fishshell.com/}.
\newblock


\bibitem[{Free Software Foundation, Inc.}(2020)]%
        {bash}
\bibfield{author}{\bibinfo{person}{{Free Software Foundation, Inc.}}}
  \bibinfo{year}{2020}\natexlab{}.
\newblock \bibinfo{title}{Bash - {GNU} {Project}}.
\newblock \bibinfo{howpublished}{https://www.gnu.org/software/bash/}.
\newblock


\bibitem[Furnas(1986)]%
        {doi:10.1145/22627.22342}
\bibfield{author}{\bibinfo{person}{G.~W. Furnas}.}
  \bibinfo{year}{1986}\natexlab{}.
\newblock \showarticletitle{Generalized fisheye views}. In
  \bibinfo{booktitle}{\emph{Proceedings of the {SIGCHI} {Conference} on {Human}
  {Factors} in {Computing} {Systems}}}. ACM, \bibinfo{pages}{16--23}.
\newblock


\bibitem[Furnas(2006)]%
        {doi:10.1145/1124772.1124921}
\bibfield{author}{\bibinfo{person}{George~W. Furnas}.}
  \bibinfo{year}{2006}\natexlab{}.
\newblock \showarticletitle{A fisheye follow-up}. In
  \bibinfo{booktitle}{\emph{Proceedings of the {SIGCHI} {Conference} on {Human}
  {Factors} in {Computing} {Systems}}}. ACM, \bibinfo{pages}{999--1008}.
\newblock


\bibitem[{GitHub, Inc.}(2024)]%
        {copilot}
\bibfield{author}{\bibinfo{person}{{GitHub, Inc.}}}
  \bibinfo{year}{2024}\natexlab{}.
\newblock \bibinfo{title}{GitHub {Copilot} \textperiodcentered{} {Your} {AI}
  pair programmer}.
\newblock \bibinfo{howpublished}{https://github.com/features/copilot}.
\newblock


\bibitem[Greenberg and Blatt(2019)]%
        {doi:10.1145/3371111}
\bibfield{author}{\bibinfo{person}{Michael Greenberg} {and}
  \bibinfo{person}{Austin~J. Blatt}.} \bibinfo{year}{2019}\natexlab{}.
\newblock \showarticletitle{Executable formal semantics for the {POSIX} shell}.
\newblock \bibinfo{journal}{\emph{Proceedings of the ACM on Programming
  Languages}}  \bibinfo{volume}{4} (\bibinfo{date}{dec 20}
  \bibinfo{year}{2019}), \bibinfo{pages}{1--30}.
\newblock


\bibitem[Greenberg et~al\mbox{.}(2021a)]%
        {doi:10.1145/3458336.3465296}
\bibfield{author}{\bibinfo{person}{Michael Greenberg},
  \bibinfo{person}{Konstantinos Kallas}, {and} \bibinfo{person}{Nikos
  Vasilakis}.} \bibinfo{year}{2021}\natexlab{a}.
\newblock \showarticletitle{The future of the shell}. In
  \bibinfo{booktitle}{\emph{Proceedings of the {Workshop} on {Hot} {Topics} in
  {Operating} {Systems}}}. ACM, \bibinfo{pages}{240--241}.
\newblock


\bibitem[Greenberg et~al\mbox{.}(2021b)]%
        {doi:10.1145/3458336.3465294}
\bibfield{author}{\bibinfo{person}{Michael Greenberg},
  \bibinfo{person}{Konstantinos Kallas}, {and} \bibinfo{person}{Nikos
  Vasilakis}.} \bibinfo{year}{2021}\natexlab{b}.
\newblock \showarticletitle{Unix shell programming}. In
  \bibinfo{booktitle}{\emph{Proceedings of the {Workshop} on {Hot} {Topics} in
  {Operating} {Systems}}}. ACM, \bibinfo{pages}{104--111}.
\newblock


\bibitem[Greenberg et~al\mbox{.}(2021c)]%
        {Greenberg2021ReportOT}
\bibfield{author}{\bibinfo{person}{Michael~R. Greenberg},
  \bibinfo{person}{Konstantinos Kallas}, \bibinfo{person}{Nikos Vasilakis},
  {and} \bibinfo{person}{Stephen Kell}.} \bibinfo{year}{2021}\natexlab{c}.
\newblock \showarticletitle{Report on the "{The} {Future} of the {Shell}"
  {Panel} at {HotOS} 2021}.
\newblock
  \bibinfo{howpublished}{https://api.semanticscholar.org/CorpusID:237605419}.
\newblock \bibinfo{journal}{\emph{ArXiv}}  \bibinfo{volume}{abs/2109.11016}
  (\bibinfo{year}{2021}).
\newblock


\bibitem[Haverbeke(2024)]%
        {codemirror}
\bibfield{author}{\bibinfo{person}{Marijn Haverbeke}.}
  \bibinfo{year}{2024}\natexlab{}.
\newblock \bibinfo{title}{CodeMirror}.
\newblock \bibinfo{howpublished}{https://codemirror.net/}.
\newblock


\bibitem[Healy(2024)]%
        {sociologists}
\bibfield{author}{\bibinfo{person}{Kieran Healy}.}
  \bibinfo{year}{2024}\natexlab{}.
\newblock \bibinfo{title}{Modern {Plain} {Text} {Computing}}.
\newblock \bibinfo{howpublished}{https://mptc.io/}.
\newblock


\bibitem[Hempel et~al\mbox{.}(2019)]%
        {doi:10.1145/3332165.3347925}
\bibfield{author}{\bibinfo{person}{Brian Hempel}, \bibinfo{person}{Justin
  Lubin}, {and} \bibinfo{person}{Ravi Chugh}.} \bibinfo{year}{2019}\natexlab{}.
\newblock \showarticletitle{Sketch-n-{Sketch}}. In
  \bibinfo{booktitle}{\emph{Proceedings of the 32nd {Annual} {ACM} {Symposium}
  on {User} {Interface} {Software} and {Technology}}}. ACM,
  \bibinfo{pages}{281--292}.
\newblock


\bibitem[Hoffswell et~al\mbox{.}(2018)]%
        {doi:10.1145/3173574.3174106}
\bibfield{author}{\bibinfo{person}{Jane Hoffswell}, \bibinfo{person}{Arvind
  Satyanarayan}, {and} \bibinfo{person}{Jeffrey Heer}.}
  \bibinfo{year}{2018}\natexlab{}.
\newblock \showarticletitle{Augmenting {Code} with {In} {Situ} {Visualizations}
  to {Aid} {Program} {Understanding}}. In \bibinfo{booktitle}{\emph{Proceedings
  of the 2018 {CHI} {Conference} on {Human} {Factors} in {Computing}
  {Systems}}}. ACM, \bibinfo{pages}{1--12}.
\newblock


\bibitem[Holen(2024)]%
        {shellcheck}
\bibfield{author}{\bibinfo{person}{Vidar Holen}.}
  \bibinfo{year}{2024}\natexlab{}.
\newblock \bibinfo{title}{ShellCheck -- shell script analysis tool}.
\newblock \bibinfo{howpublished}{https://www.shellcheck.net/}.
\newblock


\bibitem[Horowitz and Heer(2023a)]%
        {doi:10.1145/3586183.3606733}
\bibfield{author}{\bibinfo{person}{Joshua Horowitz} {and}
  \bibinfo{person}{Jeffrey Heer}.} \bibinfo{year}{2023}\natexlab{a}.
\newblock \showarticletitle{Engraft: An {API} for {Live}, {Rich}, and
  {Composable} {Programming}}. In \bibinfo{booktitle}{\emph{Proceedings of the
  36th {Annual} {ACM} {Symposium} on {User} {Interface} {Software} and
  {Technology}}}. ACM, \bibinfo{pages}{1--18}.
\newblock


\bibitem[Horowitz and Heer(2023b)]%
        {lrc}
\bibfield{author}{\bibinfo{person}{Joshua Horowitz} {and}
  \bibinfo{person}{Jeffrey Heer}.} \bibinfo{year}{2023}\natexlab{b}.
\newblock \bibinfo{title}{Live, {Rich}, and {Composable}: Qualities for
  {Programming} {Beyond} {Static} {Text}}.  (\bibinfo{year}{2023}).
\newblock
\newblock
\shownote{Presented at the 13th annual workshop on the intersection of HCI and
  PL (PLATEAU 2023)}.


\bibitem[Huang et~al\mbox{.}(2022)]%
        {doi:10.1145/3478431.3499305}
\bibfield{author}{\bibinfo{person}{Ruanqianqian~(Lisa) Huang},
  \bibinfo{person}{Kasra Ferdowsi}, \bibinfo{person}{Ana Selvaraj},
  \bibinfo{person}{Adalbert~Gerald Soosai~Raj}, {and} \bibinfo{person}{Sorin
  Lerner}.} \bibinfo{year}{2022}\natexlab{}.
\newblock \showarticletitle{Investigating the {Impact} of {Using} a {Live}
  {Programming} {Environment} in a {CS1} {Course}}. In
  \bibinfo{booktitle}{\emph{Proceedings of the 53rd {ACM} {Technical}
  {Symposium} on {Computer} {Science} {Education}}}. ACM,
  \bibinfo{pages}{495--501}.
\newblock


\bibitem[Jakubovic et~al\mbox{.}(2023)]%
        {doi:10.22152/programming-journal.org/2023/7/13}
\bibfield{author}{\bibinfo{person}{Joel Jakubovic}, \bibinfo{person}{Jonathan
  Edwards}, {and} \bibinfo{person}{Tomas Petricek}.}
  \bibinfo{year}{2023}\natexlab{}.
\newblock \showarticletitle{Technical {Dimensions} of {Programming} {Systems}}.
\newblock \bibinfo{journal}{\emph{The Art, Science, and Engineering of
  Programming}} \bibinfo{volume}{7}, \bibinfo{number}{3} (\bibinfo{date}{feb
  15} \bibinfo{year}{2023}).
\newblock


\bibitem[Kamara(2024)]%
        {explainshell}
\bibfield{author}{\bibinfo{person}{Idan Kamara}.}
  \bibinfo{year}{2024}\natexlab{}.
\newblock \bibinfo{title}{explainshell.com - match command-line arguments to
  their help text}.
\newblock \bibinfo{howpublished}{https://www.explainshell.com/}.
\newblock


\bibitem[Kang and Guo(2017)]%
        {doi:10.1145/3126594.3126632}
\bibfield{author}{\bibinfo{person}{Hyeonsu Kang} {and}
  \bibinfo{person}{Philip~J. Guo}.} \bibinfo{year}{2017}\natexlab{}.
\newblock \showarticletitle{Omnicode}. In \bibinfo{booktitle}{\emph{Proceedings
  of the 30th {Annual} {ACM} {Symposium} on {User} {Interface} {Software} and
  {Technology}}}. ACM, \bibinfo{pages}{737--745}.
\newblock


\bibitem[Kasibatla and Warth(2017)]%
        {seymour}
\bibfield{author}{\bibinfo{person}{Saketh Kasibatla} {and}
  \bibinfo{person}{Alex Warth}.} \bibinfo{year}{2017}\natexlab{}.
\newblock \bibinfo{title}{Seymour: Live {Programming} for the {Classroom}}.
  (\bibinfo{year}{2017}).
\newblock
\newblock
\shownote{Presented at the Workshop on Live Programming (LIVE) 2017}.


\bibitem[Kleppmann et~al\mbox{.}(2019)]%
        {doi:10.1145/3359591.3359737}
\bibfield{author}{\bibinfo{person}{Martin Kleppmann}, \bibinfo{person}{Adam
  Wiggins}, \bibinfo{person}{Peter van Hardenberg}, {and} \bibinfo{person}{Mark
  McGranaghan}.} \bibinfo{year}{2019}\natexlab{}.
\newblock \showarticletitle{Local-first software: you own your data, in spite
  of the cloud}. In \bibinfo{booktitle}{\emph{Proceedings of the 2019 {ACM}
  {SIGPLAN} {International} {Symposium} on {New} {Ideas}, {New} {Paradigms},
  and {Reflections} on {Programming} and {Software}}}. ACM,
  \bibinfo{pages}{154--178}.
\newblock


\bibitem[Kluyver et~al\mbox{.}(2016)]%
        {Kluyver2016JupyterN}
\bibfield{author}{\bibinfo{person}{Thomas Kluyver}, \bibinfo{person}{Benjamin
  Ragan-Kelley}, \bibinfo{person}{Fernando P{\' e}rez},
  \bibinfo{person}{Brian~E. Granger}, \bibinfo{person}{Matthias Bussonnier},
  \bibinfo{person}{Jonathan Frederic}, \bibinfo{person}{Kyle Kelley},
  \bibinfo{person}{Jessica~B. Hamrick}, \bibinfo{person}{Jason Grout},
  \bibinfo{person}{Sylvain Corlay}, \bibinfo{person}{Paul Ivanov},
  \bibinfo{person}{Dami{\' a}n Avila}, \bibinfo{person}{Safia Abdalla},
  \bibinfo{person}{Carol Willing}, {and} \bibinfo{person}{Jupyter~Development
  Team}.} \bibinfo{year}{2016}\natexlab{}.
\newblock \showarticletitle{Jupyter {Notebooks} - a publishing format for
  reproducible computational workflows}. In
  \bibinfo{booktitle}{\emph{International {Conference} on {Electronic}
  {Publishing}}}.
\newblock


\bibitem[Lerner(2020)]%
        {doi:10.1145/3313831.3376494}
\bibfield{author}{\bibinfo{person}{Sorin Lerner}.}
  \bibinfo{year}{2020}\natexlab{}.
\newblock \showarticletitle{Projection {Boxes}: On-the-fly {Reconfigurable}
  {Visualization} for {Live} {Programming}}. In
  \bibinfo{booktitle}{\emph{Proceedings of the 2020 {CHI} {Conference} on
  {Human} {Factors} in {Computing} {Systems}}}. ACM, \bibinfo{pages}{1--7}.
\newblock


\bibitem[Maci{\' a}(2024)]%
        {fsatrace}
\bibfield{author}{\bibinfo{person}{Jorge~Acereda Maci{\' a}}.}
  \bibinfo{year}{2024}\natexlab{}.
\newblock \bibinfo{title}{Filesystem {Access} {Tracer}}.
\newblock \bibinfo{howpublished}{https://github.com/jacereda/fsatrace}.
\newblock


\bibitem[Mart{\' i}(2024)]%
        {mvdan-sh}
\bibfield{author}{\bibinfo{person}{Daniel Mart{\' i}}.}
  \bibinfo{year}{2024}\natexlab{}.
\newblock \bibinfo{title}{mvdan/sh}.
\newblock \bibinfo{howpublished}{https://github.com/mvdan/sh}.
\newblock


\bibitem[Maxwell(2015)]%
        {pipecut}
\bibfield{author}{\bibinfo{person}{David~W. Maxwell}.}
  \bibinfo{year}{2015}\natexlab{}.
\newblock \bibinfo{title}{Pipecut {Project} {Home}}.
\newblock
  \bibinfo{howpublished}{https://web.archive.org/web/20220409090300/pipecut.org}.
\newblock


\bibitem[McGregor(2024)]%
        {data-journalists}
\bibfield{author}{\bibinfo{person}{Susan McGregor}.}
  \bibinfo{year}{2024}\natexlab{}.
\newblock \bibinfo{title}{Adding {Headers} and {Using} {Shell} {Scripts}}.
\newblock
  \bibinfo{howpublished}{https://data.journalism.columbia.edu/content/adding-headers-and-using-shell-scripts}.
\newblock


\bibitem[{Meta Platforms, Inc}(2024)]%
        {react}
\bibfield{author}{\bibinfo{person}{{Meta Platforms, Inc}}.}
  \bibinfo{year}{2024}\natexlab{}.
\newblock \bibinfo{title}{React}.
\newblock \bibinfo{howpublished}{https://react.dev/}.
\newblock


\bibitem[{Microsoft Corporation}(2024)]%
        {powershell}
\bibfield{author}{\bibinfo{person}{{Microsoft Corporation}}.}
  \bibinfo{year}{2024}\natexlab{}.
\newblock \bibinfo{title}{How to {Debug} {Scripts} in {Windows} {PowerShell}
  {ISE}}.
\newblock
  \bibinfo{howpublished}{https://learn.microsoft.com/en-us/powershell/scripting/windows-powershell/ise/how-to-debug-scripts-in-windows-powershell-ise}.
\newblock


\bibitem[Morgan(2024)]%
        {murex}
\bibfield{author}{\bibinfo{person}{Laurence Morgan}.}
  \bibinfo{year}{2024}\natexlab{}.
\newblock \bibinfo{title}{Murex}.
\newblock \bibinfo{howpublished}{https://murex.rocks/}.
\newblock


\bibitem[Nardi(1993)]%
        {Nardi1993-ln}
\bibfield{author}{\bibinfo{person}{Bonnie~A Nardi}.}
  \bibinfo{year}{1993}\natexlab{}.
\newblock \bibinfo{booktitle}{\emph{A small matter of programming}}.
\newblock \bibinfo{publisher}{MIT Press}, \bibinfo{address}{London, England}.
\newblock


\bibitem[Norman and Draper(1986)]%
        {doi:10.1201/b15703}
\bibfield{author}{\bibinfo{person}{Donald~A. Norman} {and}
  \bibinfo{person}{Stephen~W. Draper}.} \bibinfo{year}{1986}\natexlab{}.
\newblock \bibinfo{booktitle}{\emph{User {Centered} {System} {Design}}}.
\newblock \bibinfo{publisher}{CRC Press}.
\newblock


\bibitem[{Nushell Project Developers}(2024)]%
        {nushell}
\bibfield{author}{\bibinfo{person}{{Nushell Project Developers}}.}
  \bibinfo{year}{2024}\natexlab{}.
\newblock \bibinfo{title}{Nushell}.
\newblock \bibinfo{howpublished}{https://www.nushell.sh/}.
\newblock


\bibitem[{Observable Inc.}(2022)]%
        {observable}
\bibfield{author}{\bibinfo{person}{{Observable Inc.}}}
  \bibinfo{year}{2022}\natexlab{}.
\newblock \bibinfo{title}{Observable - {Explore}, analyze, and explain data.
  {As} a team.}
\newblock \bibinfo{howpublished}{https://observablehq.com/}.
\newblock


\bibitem[Omar et~al\mbox{.}(2021)]%
        {doi:10.1145/3453483.3454059}
\bibfield{author}{\bibinfo{person}{Cyrus Omar}, \bibinfo{person}{David Moon},
  \bibinfo{person}{Andrew Blinn}, \bibinfo{person}{Ian Voysey},
  \bibinfo{person}{Nick Collins}, {and} \bibinfo{person}{Ravi Chugh}.}
  \bibinfo{year}{2021}\natexlab{}.
\newblock \showarticletitle{Filling typed holes with live {GUIs}}. In
  \bibinfo{booktitle}{\emph{Proceedings of the 42nd {ACM} {SIGPLAN}
  {International} {Conference} on {Programming} {Language} {Design} and
  {Implementation}}}. ACM, \bibinfo{pages}{511--525}.
\newblock


\bibitem[{OpenAI, L.L.C.}(2024)]%
        {chatgpt}
\bibfield{author}{\bibinfo{person}{{OpenAI, L.L.C.}}}
  \bibinfo{year}{2024}\natexlab{}.
\newblock \bibinfo{title}{ChatGPT}.
\newblock \bibinfo{howpublished}{https://chatgpt.com/}.
\newblock


\bibitem[Pendry and McKusick(1995)]%
        {pendry1995union}
\bibfield{author}{\bibinfo{person}{Jan-Simon Pendry} {and}
  \bibinfo{person}{Marshall~Kirk McKusick}.} \bibinfo{year}{1995}\natexlab{}.
\newblock \showarticletitle{Union {Mounts} in 4.4 {BSD}-{Lite}}. In
  \bibinfo{booktitle}{\emph{USENIX 1995 {Technical} {Conference}}}.
\newblock


\bibitem[{Postman, Inc.}(2024)]%
        {postman}
\bibfield{author}{\bibinfo{person}{{Postman, Inc.}}}
  \bibinfo{year}{2024}\natexlab{}.
\newblock \bibinfo{title}{Postman {API} {Platform}}.
\newblock \bibinfo{howpublished}{https://www.postman.com/}.
\newblock


\bibitem[Pothier et~al\mbox{.}(2007)]%
        {doi:10.1145/1297105.1297067}
\bibfield{author}{\bibinfo{person}{Guillaume Pothier}, \bibinfo{person}{{\'
  E}ric Tanter}, {and} \bibinfo{person}{Jos{\' e} Piquer}.}
  \bibinfo{year}{2007}\natexlab{}.
\newblock \showarticletitle{Scalable omniscient debugging}.
\newblock \bibinfo{journal}{\emph{ACM SIGPLAN Notices}} \bibinfo{volume}{42},
  \bibinfo{number}{10} (\bibinfo{date}{oct 21} \bibinfo{year}{2007}),
  \bibinfo{pages}{535--552}.
\newblock


\bibitem[Radek~Podgorny(2024)]%
        {unionfs-fuse}
\bibfield{author}{\bibinfo{person}{Bernd~Schubert Radek~Podgorny}.}
  \bibinfo{year}{2024}\natexlab{}.
\newblock \bibinfo{title}{unionfs-fuse}.
\newblock \bibinfo{howpublished}{https://github.com/rpodgorny/unionfs-fuse}.
\newblock


\bibitem[Rauch et~al\mbox{.}(2019)]%
        {doi:10.22152/programming-journal.org/2019/3/9}
\bibfield{author}{\bibinfo{person}{David Rauch}, \bibinfo{person}{Patrick
  Rein}, \bibinfo{person}{Stefan Ramson}, \bibinfo{person}{Jens Lincke}, {and}
  \bibinfo{person}{Robert Hirschfeld}.} \bibinfo{year}{2019}\natexlab{}.
\newblock \showarticletitle{Babylonian-style {Programming}: Design and
  {Implementation} of an {Integration} of {Live} {Examples} into
  {General}-purpose {Source} {Code}}.
\newblock \bibinfo{journal}{\emph{The Art, Science, and Engineering of
  Programming}} \bibinfo{volume}{3}, \bibinfo{number}{3} (\bibinfo{date}{feb 1}
  \bibinfo{year}{2019}).
\newblock


\bibitem[Ritchie(1980)]%
        {doi:10.1007/3-540-09745-7_2}
\bibfield{author}{\bibinfo{person}{Dennis~M. Ritchie}.}
  \bibinfo{year}{1980}\natexlab{}.
\newblock \bibinfo{booktitle}{\emph{The evolution of the unix time-sharing
  system}}.
\newblock \bibinfo{publisher}{Springer Berlin Heidelberg},
  \bibinfo{pages}{25--35}.
\newblock


\bibitem[Ritchie and Thompson(1974)]%
        {doi:10.1145/361011.361061}
\bibfield{author}{\bibinfo{person}{Dennis~M. Ritchie} {and}
  \bibinfo{person}{Ken Thompson}.} \bibinfo{year}{1974}\natexlab{}.
\newblock \showarticletitle{The {UNIX} time-sharing system}.
\newblock \bibinfo{journal}{\emph{Commun. ACM}} \bibinfo{volume}{17},
  \bibinfo{number}{7} (\bibinfo{date}{7} \bibinfo{year}{1974}),
  \bibinfo{pages}{365--375}.
\newblock


\bibitem[Sul{\' i}r et~al\mbox{.}(2023)]%
        {doi:10.1145/3593434.3593501}
\bibfield{author}{\bibinfo{person}{Mat{\' u}{\v s} Sul{\' i}r},
  \bibinfo{person}{Sergej Chodarev}, {and} \bibinfo{person}{Milan Nos{\' a}{\v
  l}}.} \bibinfo{year}{2023}\natexlab{}.
\newblock \showarticletitle{Outside the {Sandbox}: A {Study} of
  {Input}/{Output} {Methods} in {Java}}. In
  \bibinfo{booktitle}{\emph{Proceedings of the 27th {International}
  {Conference} on {Evaluation} and {Assessment} in {Software} {Engineering}}}.
  ACM, \bibinfo{pages}{253--258}.
\newblock


\bibitem[{The PaSh Authors}(2024)]%
        {try}
\bibfield{author}{\bibinfo{person}{{The PaSh Authors}}.}
  \bibinfo{year}{2024}\natexlab{}.
\newblock \bibinfo{title}{binpash/try}.
\newblock \bibinfo{howpublished}{https://github.com/binpash/try}.
\newblock


\bibitem[{University of Cambridge Information Services}(2024)]%
        {scientists}
\bibfield{author}{\bibinfo{person}{{University of Cambridge Information
  Services}}.} \bibinfo{year}{2024}\natexlab{}.
\newblock \bibinfo{title}{Unix: Simple {Shell} {Scripting} for {Scientists}}.
\newblock
  \bibinfo{howpublished}{https://training.cam.ac.uk/ucs/course/ucs-scriptsci}.
\newblock


\bibitem[Victor(2006)]%
        {magic-ink}
\bibfield{author}{\bibinfo{person}{Bret Victor}.}
  \bibinfo{year}{2006}\natexlab{}.
\newblock \bibinfo{title}{Magic {Ink}: Information {Software} and the
  {Graphical} {Interface}}.
\newblock \bibinfo{howpublished}{https://worrydream.com/MagicInk/}.
\newblock


\bibitem[{xonsh developers}(2024)]%
        {xonsh}
\bibfield{author}{\bibinfo{person}{{xonsh developers}}.}
  \bibinfo{year}{2024}\natexlab{}.
\newblock \bibinfo{title}{The {Xonsh} {Shell} --- {Python}-powered shell}.
\newblock \bibinfo{howpublished}{https://xon.sh/}.
\newblock


\end{thebibliography}

\appendix
\section{Study tasks}
\label{sec:study-tasks}

Below are the written prompts for the four tasks in our study, including any starter code that was provided.

\subsection{\textbf{titles}}
\label{sec:titles}

\begin{quote}
The ``docs'' directory contains a few speeches. Each lists its title
as its first line. Write a script that renames each file to its
title.

\end{quote}

(Participants discovered that some of the titles contained spaces, which could cause trouble unless they were appropriately quoted.)

\subsection{\textbf{users}}
\label{sec:users}

\begin{quote}
We're running a web service, and need to respond to some users'
requests to delete their accounts. We have a list of usernames in a
file called users-to-delete.txt. Write a script that deletes the
user folders for each user in that file.

A code assistant generated the following code. Is it correct?

\begin{minted}{bash}
cat users-to-delete.txt | while read LINE; do
  rm -rf "users/$LINE"
done
\end{minted}

\end{quote}

(Participants discovered that \texttt{users-to-delete.txt} contained a spurious blank line, which the above script incorrectly interprets as a command to delete the entire \texttt{users} folder.)

\subsection{\textbf{log}}
\label{sec:log}

\begin{quote}
What does this script do?

\begin{minted}{bash}
cat launchd.log |
  egrep 'Sandbox restriction$' |
  sed -n 's/.*requestor = \(.*\)\[.*/\1/p' |
  sort |
  uniq -c
\end{minted}

\end{quote}

\subsection{\textbf{lines}}
\label{sec:lines}

\begin{quote}
There are some files in the \texttt{files} directory. Write a script that
computes the product of their line counts. (Ex: If there were two
files, one with 10 lines and one with 20, the product would be
200.)

The starter code below counts lines, but doesn't do any
multiplying.

\begin{minted}{bash}
product=1
for f in files/*; do
  wc -l < $f
done
\end{minted}

\end{quote}

\end{document}